\UseRawInputEncoding

\def\PRL{{ Phys. Rev. Lett.\ }\/}
\def\PRB{{ Phys. Rev. B\ }\/}

\def\be{\begin {equation}}
\def\ee{\end {equation}}
\def\ber{\begin {eqnarray}}
\def\eer{\end {eqnarray}}
\def\bers{\begin {eqnarray*}}
\def\eers{\end {eqnarray*}}

\newcommand{\SPhide}[1]{{}}

\newcommand{\SPEDITOKAY}[2]{{}{\textcolor{black}{#2}}}

\newcommand{\SPHIDE}[1]{{}}

\newcommand{\CMEDITOKAY}[2]{{}{\textcolor{black}{#2}}}

\newcommand{\AAA}[1]{{\textcolor{black}{ #1}}}

\makeatletter

\newcommand{\Rmnum}[1]{\expandafter\@slowromancap\romannumeral #1@}
\makeatother

\makeatletter
\newcommand*\env@matrix[1][*\c@MaxMatrixCols c]{%
  \hskip -\arraycolsep
  \let\@ifnextchar\new@ifnextchar
  \array{#1}}
\makeatother

\documentclass[aps,prb,showpacs,superscriptaddress,a4paper,twocolumn]{revtex4-1}
\usepackage[usenames, dvipsnames]{color}
\usepackage[normalem]{ulem}
\usepackage{amsmath}
\usepackage{float}
\usepackage{color}

\usepackage[normalem]{ulem}
\usepackage{soul}
\usepackage{amssymb}

\usepackage{amsfonts}
\usepackage{amssymb}
\usepackage{graphicx}
\usepackage{forest}
\definecolor{ginger}{rgb}{0.69, 0.4, 0.0}
\usepackage{appendix}

\begin {document}

\title{Symmetry protection and giant Fermi arcs from higher-fold fermions in binary, ternary and quaternary compounds} 

\author{Chanchal K. Barman}
\thanks{These two authors have contributed equally to this work}
\affiliation{Department of Physics, Indian Institute of Technology, Bombay, Powai, Mumbai 400076, India}

\author{Chiranjit Mondal}
\thanks{These two authors have contributed equally to this work}
\affiliation{Discipline of Metallurgy Engineering and Materials Science, IIT Indore, Simrol, Indore 453552, India}

\author{Sumiran Pujari}
\affiliation{Department of Physics, Indian Institute of Technology, Bombay, Powai, Mumbai 400076, India}

\author{Biswarup Pathak}
\email{biswarup@iiti.ac.in }
\affiliation{Discipline of Metallurgy Engineering and Materials Science, IIT Indore, Simrol, Indore 453552, India}
\affiliation{Discipline of Chemistry, School of Basic Sciences, IIT Indore, Simrol, Indore 453552, India}

\author{Aftab Alam}
\email{aftab@iitb.ac.in}
\affiliation{Department of Physics, Indian Institute of Technology, Bombay, Powai, Mumbai 400076, India}

\date{\today}

\begin{abstract}
Higher-fold chiral fermions that go beyond two-fold Weyl fermions have recently been reported in crystalline systems. \textcolor{black}{Here, we focus on such excitations in several binary, ternary and quaternary alloys/compounds with CoGe, BiSbPt and KMgBO$_3$ as the representative examples that belong to the crystal space group (SG) 198}. We found distinct three-fold, four-fold and six-fold chiral fermions in the bulk via Density Functional computations. We provide general symmetry arguments for the protection of these degeneracies with special emphasis on the four-fold fermions. Our surface spectra simulations show that the size of Fermi arcs resulting from these chiral fermions are large, robust and untouched from the bulk states due to the near absence of trivial bulk Fermi pockets. All these features make these systems -- especially CoGe and KMgBO$_3$ -- promising topological semimetal candidates to realize higher-fold fermions in future photo-emission and transport experiments. 

\end{abstract}

\maketitle


\section{Introduction} The discovery of chiral fermions in solid state quantum materials has kick-started a burst of activity in condensed matter physics. A methodological approach towards the understanding and search of new topological semimetals is to examine how the symmetries of a material enforce or ``symmetry-protect" degenerate multi-fold band-crossing points.\citep{Rappe2012,Andrei2012} These new quasiparticles\citep{Bradlyn2016,PBPal2011,Alexey2015,Ashvin2018} in the solid state\cite{CKB2019,dDirac2016,CM2019,MoP2017,NLS-1,NLS-2,NLS-3,NLS-4,RhSi2017,FeSi2018,SCZHANG2017}
may not even have elementary particle counterparts.

\begin{figure}[h]
\centering
\includegraphics[width=0.9\linewidth]{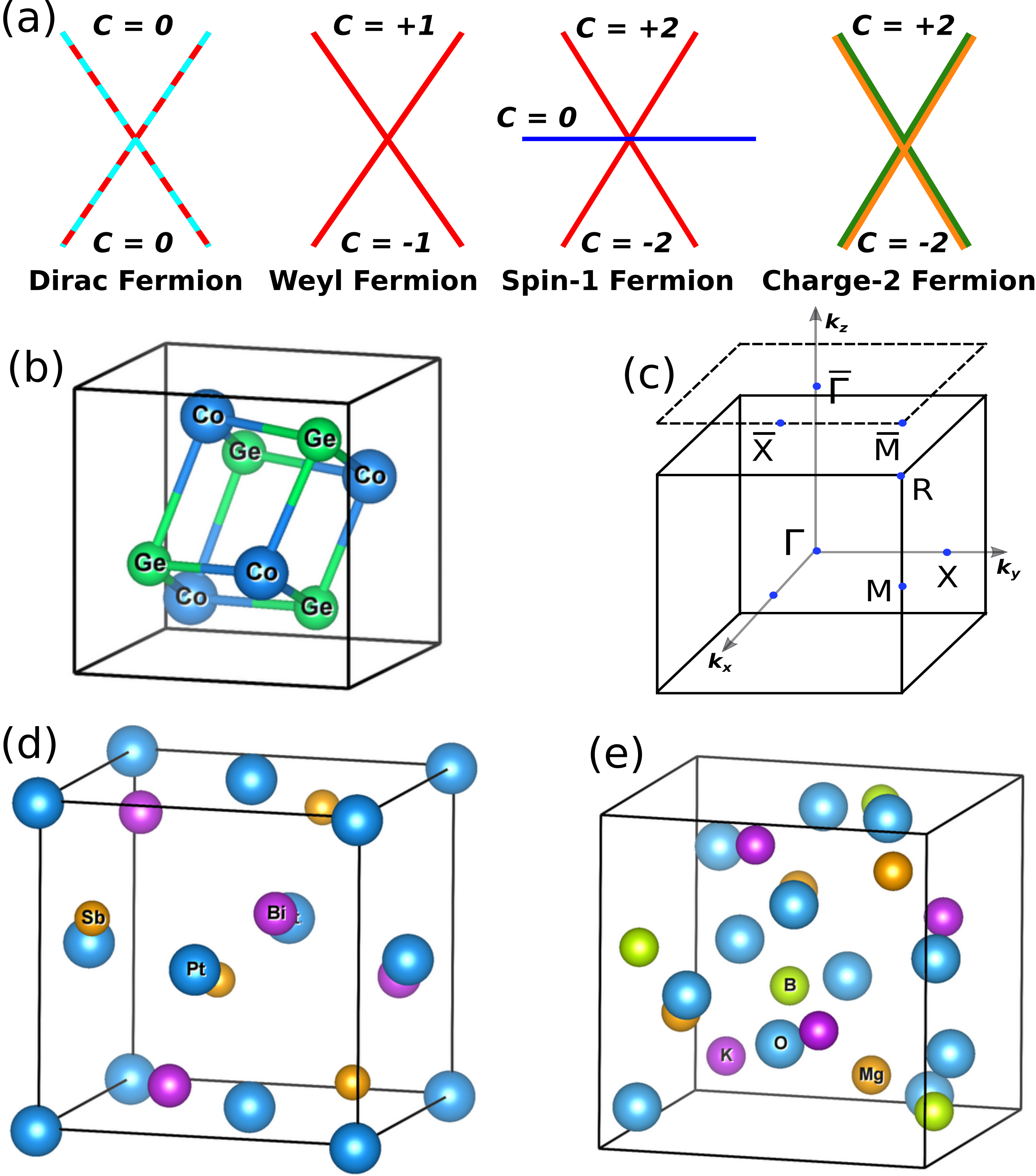}
\caption{(Color online) (a) Schematic band structure of Dirac, Weyl, Spin-1 and Charge-2 fermions. (b) Crystal Structure of CoGe (space group P2$_1$3). (c) Bulk Brillouin zone (BZ) and (001) surface BZ (represented by dashed square). Crystal structure of (d) ternary BiSbPt and (e) quaternary KMgBO$_3$. }
\label{fig1}
\end{figure}

Some of the new, unexpected quasiparticle excitations predicted recently are spin-1,\cite{RhSi2017,FeSi2018,SCZHANG2017,TZHANG2018} charge-2 Dirac,\cite{SCZHANG2017,TZHANG2018} and spin-$\frac{3}{2}$\cite{SCZHANG2017} chiral fermions. The well-known two-fold Weyl chiral fermions can be present in the absence of either inversion ($\mathcal{I}$) or time reversal ($\mathcal{T}$) symmetry in three dimensional (3D) crystals. They are characterized by non-zero topological charges called Chern numbers (C=$\pm$1).\citep{Alexey2015,Ashvin2018}
These Weyl fermions can be described by an effective spin-$\frac{1}{2}$ Hamiltonian $H \propto \hbar \: \delta \vec{\textbf {k}}\cdot\vec{\mathbf{\sigma}}$ at lowest order. $\delta\vec{\textbf {k}}$ is small deviations from Weyl node in momentum space. $\vec{\mathbf{\sigma}} \equiv \{\sigma_x, \sigma_y, \sigma_z \}$ are the 2$\times$2 Pauli matrices. 
However, certain crystal symmetries can also protect spin-1 or spin-$\frac{3}{2}$ chiral fermions\citep{SCZHANG2017,TZHANG2018} at high symmetry points in the crystal momentum space. They are three-fold and four-fold respectively. Their effective low-energy Hamiltonians are $H \propto \hbar \: \delta \vec{\textbf {k}}\cdot\vec{\mathbf{L}}$, where $L_i$'s are (3$\times$3) spin-1 and (4$\times$4) spin-$\frac{3}{2}$ rotation generators respectively.
\SPEDITOKAY{The Weyl nodes and Chern numbers for spin-1 chiral fermions are derived from the Hamiltonian looks like, $H_2 \propto (\hbar/2)\delta \vec{\textbf {k}}\cdot\vec{\mathbf{L}}$.}{} 
The low energy dispersions of these ``multi-Weyl" systems follow from the corresponding model Hamiltonians, 
e.g. spin-1 fermions possess a combination of a Dirac-type linear band crossing and a flat band, as shown in Fig.~\ref{fig1}(a), with C=$\pm2$ and $0$ respectively.

Additionally, two identical copies of spin-$\frac{1}{2}$ Weyl nodes can also be symmetry-protected.\cite{FeSi2018,TZHANG2018} This leads to $C=\pm2$ with four-fold degeneracy. The effective  Hamiltonian for such a multi-Weyl node\cite{FeSi2018,TZHANG2018} can be described as $H \propto \hbar \: \delta \vec{\textbf {k}}\cdot\vec{\mathbf{\sigma}}\otimes \mathbb{I}_{2\times2}$.  They have been named as charge-2 Dirac nodes. Fig.~\ref{fig1}(a) shows schematic diagrams of low energy dispersions for Dirac, Weyl, spin-1 and charge-2 Dirac nodes. 
The symmetry-protected band-crossings which carry C=$\pm2$ are referred to as double Weyl nodes. 
These band-crossings are topologically robust under infinitesimal changes of the Hamiltonian parameters\cite{chern_footnote} and lead to quite interesting phenomena.\cite{Ashvin2018}

{\par} In the search for such multi-Weyl systems, there have been a few studies on binary transition metal silicides with SG 198 which are predicted to be double Weyl semimetals.\citep{RhSi2017,FeSi2018,SCZHANG2017,TZHANG2018,CoSiARPES2019,CoSi2019,CoSi2019ARPES,AlPt2018} \CMEDITOKAY{}{ \textcolor{black}{Here, we study several binary, ternary and quaternary alloys with CoGe, BiSbPt and KMgBO$_3$ as the representative case respectively}}. We provide a detailed analysis including \emph{ab initio} simulations of bulk and surface excitations and symmetry protection arguments for the various multi-fold degeneracies. \textcolor{black}{Unlike previous reports which were geared towards binary systems, our symmetry arguments are quite general in the spirit of Kramer's theorem and are independent of the composition of the constituent elements.} We performed \emph{ab initio} electronic structure calculations using Vienna Ab-initio Simulation Package (VASP)\cite{GKRESSE1993,JOUBERT1999} with Perdew-Burke-Ernzerhof (PBE)\cite{JOUBERT1999} exchange correlation. Chern numbers were calculated using Wannier charge center (WCC) evolution of Maximally Localized wannier functions (MLWF)\cite{MARZARI1997,SOUZA2001,VANDERBIT} from wannier90.\cite{w90} Surface spectra and Fermi arcs were simulated using iterative Green's function method.\cite{DHLEE_I,DHLEE_II,SANCHO1985} Further information on computational details can be found in the supplementary (SI).\cite{supp}


\section { Crystal Structures}
The crystal structure and the corresponding Brillouin zone (BZ) for CoGe are shown in Fig.~\ref{fig1}(b,c). CoGe crystallizes in cubic structure with \CMEDITOKAY{space group}{SG} P2$_1$3 under high pressures.\cite{CoGeExpt,CoGefootnote} The primitive cell contains four formula units with both Co and Ge atoms lying on three-fold axes occupying the same Wyckoff sites 4a ($x,x, x$).
 The internal co-ordinates are $x_{\text{Co}}$ = 0.1359 and $x_{\text{Ge}}$ = 0.8393. The theoretically optimized lattice parameter of CoGe is found to be 4.64 \AA which matches fairly well with the experimental value, 4.637 \AA.\cite{CoGeExpt}
Figure \ref{fig1}(d,e) shows the crystal structure of BiSbPt and KMgBO$_3$ compounds.
Similar to binary CoGe, the primitive cell of ternary BiSbPt contains four formula units with Bi, Sb and Pt occupying 4a($x,x, x$) Wyckoff sites where $x_{\text{Bi}} = 0.629$, $x_{\text{Sb}} = 0.373$ and $x_{\text{Pt}} = 0.990$. The optimized lattice parameter for BiSbPt is found to be 6.69 \AA.
In KMgBO$_3$, the K, Mg, and B atoms are located in one crystallographic position $4a(x,x,x)$, while the O atoms sit on a different Wyckoff site $12b(y^1,y^2,y^3)$, where $x_{\text{K}}=0.1333$, $x_{\text{Mg}}=0.8552$ , $x_{\text{B}}=0.4076$ , $y^1_{\text{O}}=0.4181$, $y^2_{O}=0.2572$ and $y^3_{O}=0.5405$. The optimized lattice parameter for KMgBO$_3$ is found to be 6.89\AA  which is in fair agreement with the experimental value 6.83451 \AA.\cite{KMgBO3-2010}

 

\section{Symmetry Arguments}
The crystal structure \CMEDITOKAY{}{of these systems} has tetrahedral (T$_4$) point group symmetry with the following information germane to our analysis.\cite{CracknellBook} The point group has three generators at $\Gamma$ point: two screws, $S_{2z}=\{C_{2z}|\frac{1}{2},0,\frac{1}{2}\}$, $S_{2y}=\{C_{2y}|0,\frac{1}{2},\frac{1}{2}\}$ and a three-fold rotation $S_3=\{C^{+}_{3,111}|0,0,0\}$. They satisfy $S_{2z} S_3 = S_3 S_{2y}$ and $S_3 S_{2z} S_{2y} = S_{2y} S_3$. Due to $S_3$, the third screw symmetry $S_{2x}=\{C_{2x}|\frac{1}{2},\frac{1}{2},0\}$ is also present. On the otherhand at the $R$ point, the three generators are $S_{2x}=\{ C_{2x}|\frac{1}{2},\frac{3}{2},0\}$, $S_{2y}=\{C_{2y}|0,\frac{3}{2},\frac{1}{2}\}$, and $S_{3}=\{C^{-1}_{3,111}|0,1,0\}$. They satisfy $S_{2x} S_3 = S_3 S_{2y}$ and $S_3 S_{2x} S_{2y} = S_{2y} S_3$. We also keep time reversal symmetry and thus will focus on the time reversal invariant momenta in the BZ.

We start with the spinless case for which time reversal operator ($\mathcal{T}$) squares to identity ($\mathbb{I}$). \SPEDITOKAY{}{\textcolor{black}{This case is relevant for systems composed of light elements with weak spin-orbit coupling (e.g. KMgBO$_3$), as well as for phonon spectra\cite{TZHANG2018} for this crystal structure.}} At the $\Gamma$-point, the electronic structure can potentially show a three-fold band degeneracy. However, the $\Gamma$ point symmetries do not necessarily imply three-fold degeneracies. For a three-fold degeneracy, the two screw symmetries $S_{2y}$ and $S_{2z}$ should \emph{commute} and square to $\mathbb{I}$ as is the case at $\Gamma$, as well as $S_{3}$ should act non-trivially ($S_3 |\psi\rangle \neq |\psi\rangle$ where $|\psi\rangle$ is a simultaneous eigenstate of $S_{2y}$ and $S_{2z}$; see supplementary Sec. I.C of Ref. \onlinecite{Bradlyn2016}). It turns out that there can also be two-fold degeneracies or one-fold states at $\Gamma$ point consistent with the symmetries if $S_3$ is trivial.

The symmetry properties at $R$ point are crucially different. At this point, the two screws $S_{2x}$ and $S_{2y}$ now \emph{anticommute} and square to $-\mathbb{I}$, and hence the previous three-fold degeneracy argument does not apply anymore. Ref.~\onlinecite{TZHANG2018} offered an intuition that the degeneracy at $R$ point has to be even dimensional with a lower bound of four.\cite{zhang_footnote} From our analysis, we shall show that it has to be even with an upper bound of four in presence of $S_3$.

 Firstly, we can get a two-fold degeneracy using the anticommutation of the screws: $|\psi\rangle$ and $S_{2x} |\psi\rangle$ are distinct eigenstates under $S_{2y}$, say with eigenvalues of $i$ and $-i$ respectively without loss of generality. We can get a further two-fold degeneracy due to $S_{2z} S_3 = S_3 S_{2y}$: $S_3 |\psi\rangle$ and $S_{2y} S_3 |\psi\rangle$ are distinct eigenstates now under $S_{2x}$ with eigenvalues $i$ and $-i$ respectively. If $S_3$ is non-trivial\cite{s3_footnote} and takes us out of the subspace of $|\psi\rangle$ and $S_{2x} |\psi\rangle$, i.e. minimally $\langle \psi | S_3 \psi \rangle=0$, then mutual orthogonality of the two pairs is ensured.\cite{verbose} Time reversal (effectively complex conjugation) does not generate any new states for spinless electrons. Since we have accounted for all the symmetries present at $R$, we can at most get a symmetry-protected four-fold degeneracy and \emph{no higher}. Combining with the argument of Ref.~\onlinecite{TZHANG2018}, we arrive at an \emph{exactly} four-fold node protected by symmetries. 

\begin{figure*}[t]
\centering
\includegraphics[width=0.9\linewidth]{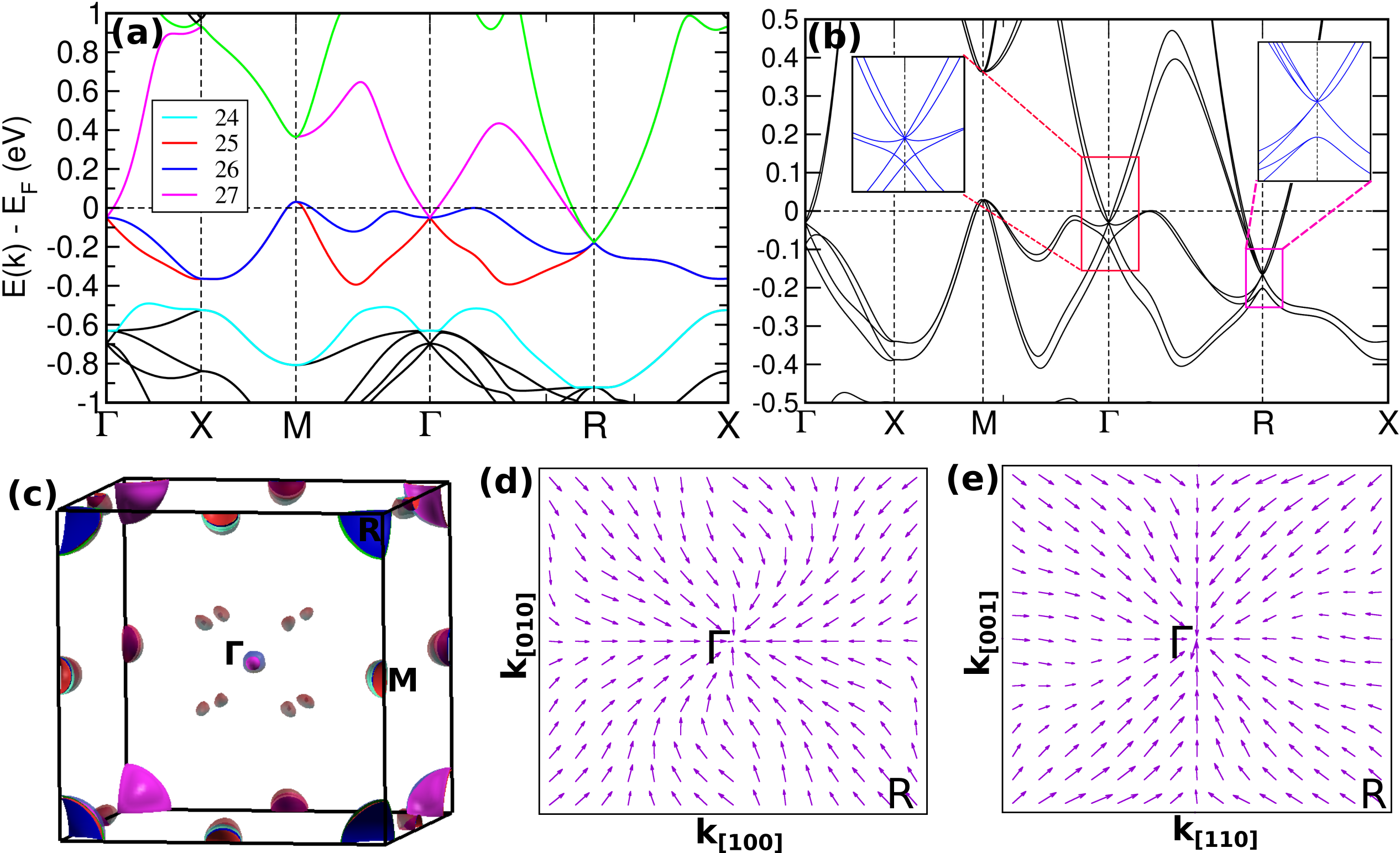}
\caption{(Color online) (a,b) Band structure of CoGe without (left) and with (right) spin-orbit interaction (SOI). The various degeneracies at the nodal points are protected by non-symmorphic screw and three-fold rotation symmetries of SG 198 and time-reversal symmetry. (c) 3D Fermi surface at isolevel E$_F$ with SOI. (d,e) In-plane Berry Curvature plotted on k$_z$=0 (left) and k$_x$=k$_y$ (right) plane highlighting its flows between $R$ and $\Gamma$ points in agreement with the sign of the topological charges in the presence of SOI. }
\label{fig3}
\end{figure*}

Going to the spinfull case for which $\mathcal{T}^2=-\mathbb{I}$,  the Kramer's degeneracies are lifted throughout the zone except at the time-reversal invariant momenta in presence of spin-orbit interaction (SOI) because the crystal does not possess space-inversion symmetry.
Adding the spin \SPEDITOKAY{degeneracy}{quantum number} to a potential three-fold spinless degeneracy at $\Gamma$, we would like to understand what happens to the six \SPEDITOKAY{fold degenerate}{} states under SOI. It turns out that they can not give rise to a six-fold degeneracy, but at least have to split into two nodal points with four-fold degenerate and two-fold degenerate states. This is because only a four-fold degeneracy can at most be protected by $\Gamma$ point symmetries. The reason for this is that now the screws $S_{2y}$ and $S_{2z}$ anticommute (and square to $-\mathbb{I}$) at $\Gamma$ point instead of $R$ point for the spinless case.\cite{2pirot} So we can again get a four-fold degeneracy by the argument previously made for the spinless case at $R$ point. However, for spinfull case, time reversal could potentially generate new eigenstates. But, \emph{mutual} orthogonality of $S_{2y}$ and $S_{2z}$ eigenstates and their time-reversed partners is \emph{not} ensured due to imaginary eigenvalues under the screws.\cite{verbose}  Thus, we can only conclude a four-fold degeneracy and no higher. This completes the splitting argument. 
Also, a single-fold spinless band at $\Gamma$ (if $S_3$ is trivial) will give rise to Kramer's two-fold degeneracy in the spinfull case. Similarly, a four-fold spinfull degeneracy arising from a two-fold spinless degeneracy is also consistent with the symmetries.
On the other hand, at $R$ point there can be six-fold degeneracies.\cite{Bradlyn2016}

To explain the spinfull four-fold degeneracy at the $\Gamma$ point for binary systems, an alternate ``top-down" argument was given in Ref. \onlinecite{RhSi2017}. Chang \emph{et al} started with a eight dimensional representation of the Hamiltonian after making (minimal) assumptions on the nature of the orbitals in the unit cell. They then wrote down the distinct symmetry allowed ``mass" terms in the $\mathbf{k}\cdot\mathbf{p}$ Hamiltonian based on the procedure laid down in Ref. \onlinecite{Wieder_Kane} to reduce down to a four-fold degeneracy. Our arguments \cite{verbose} above are rather ``bottom-up" and purely based on symmetries of the SG. On the other hand, comparing with the arguments of Ref. \onlinecite{Bradlyn2016} for the case of commuting screws, we have paid attention to the interplay of $S_3$ symmetry of SG 198 with anticommuting screws which forbids any degeneracies higher than four-fold (and \emph{only} four-fold for spinless case at $R$ point). In particular, our arguments also predict that systems beyond the binary class, e.g. ternaries and quaternaries in SG 198 will also host these four-fold degeneracies.

We finally note that the four-fold degeneracies at $R$ point have charge-2 Dirac nodal character.
This is ensured because of the presence of two-fold line degeneracies along
$R$-$X$ and $M$-$X$ directions (in fact, the whole $k_x=\pi$ and symmetry-related planes).
Such additional symmetry protection obtains from a product of time reversal and 
screw symmetries (e.g. $\mathcal{T} S_{2x}$)\cite{verbose} 
leading to Kramer's-like two-fold degeneracies. Four-fold degeneracies at $\Gamma$ point are not so constrained and therefore generically have spin-$\frac{3}{2}$ character instead.


\begin{figure*}[t]
\centering
\includegraphics[width=0.8\linewidth]{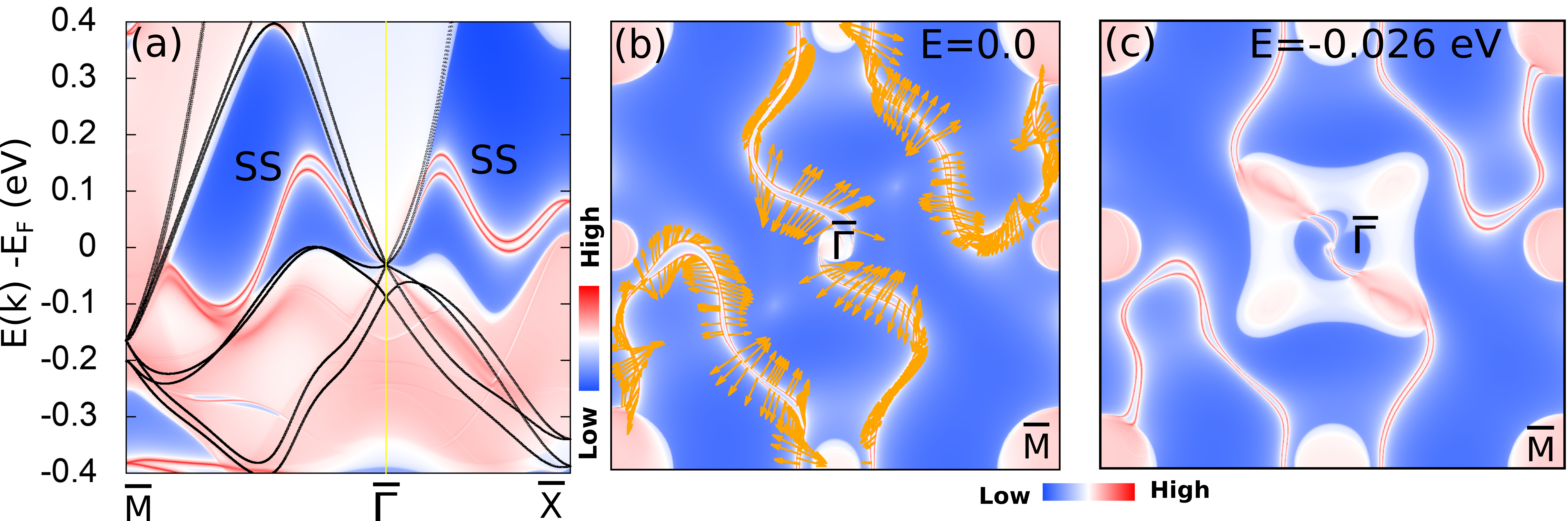}
\caption{(Color online) (a) Surface spectrum of CoGe at (001) surface in presence of SOI. Surface states are marked by SS. Superimposed bulk band structure along $R$-$\Gamma$-X are represented by black lines. \textcolor{black}{(b,c) Fermi arc contour at E$_F$ and  E$_F -0.026$ eV (spin-3/2 Weyl node).} (b) also shows the spin-momentum locked spin texture (orange arrows).}
\label{fig4}
\end{figure*}

\section{Results and Discussion}
\subsection{Binary Compound (CoGe)}
\par {\it Bulk Excitations}:\
In figure~\ref{fig3}(a) \SPEDITOKAY{}{we briefly look at the spinless} electronic band structure of CoGe \SPEDITOKAY{in the absence of SOI}{by suppressing SOI}. 
Different colored lines in Fig.~\ref{fig3}(a) indicate band index (24 to 27). 
At $\Gamma$, we see a three-fold or spin-1 degeneracy as discussed earlier. There are also two-fold degeneracies and one-fold states at $\Gamma$ at other energies (not highlighted). 
The computed Chern number for 25$^{th}$ to 27$^{th}$ bands at $\Gamma$ point are found to be C(25)=$-$2, C(26)=0 and C(27)=+2 respectively. 
On the other hand at $R$ point, we find \emph{only} four-fold degeneracies in line with the symmetry arguments. One such 4-fold degeneracy with charge-2 nodal is highlighted in Fig.~\ref{fig3}(a). The computed Chern number at this four-fold degenerate node is $+2$. Hence, the total Chern number is zero in the entire zone in accordance with the Nielsen-Ninomiya theorem.\citep{nogo} \textcolor{black}{These observations are also pertinent to the weak-SOI case of KMgBO$_3$ to be discussed later (Fig.~\ref{fig6}(a)).}

\begin{figure}[b]
\centering
\includegraphics[width=\linewidth]{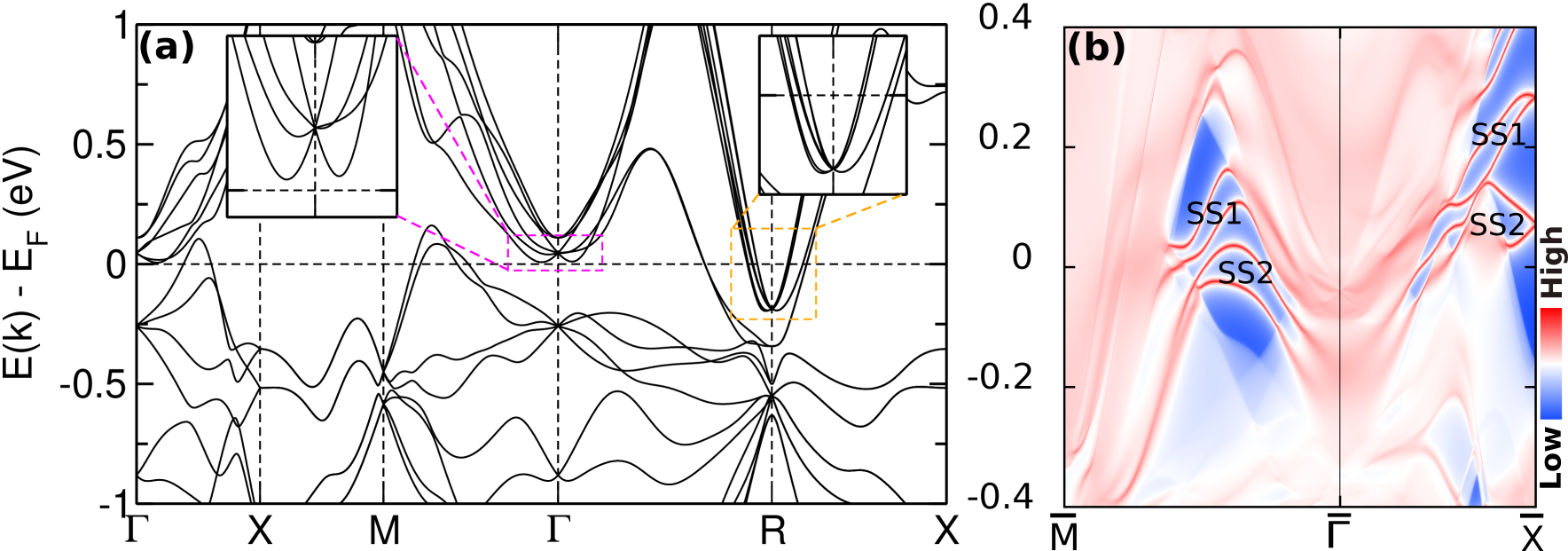}
\caption{(Color online) \textcolor{black}{For BiSbPt with SOI: (a) Electronic band structure, and (b) surface spectrum at (001) surface. Surface states are marked by SS. Inset in (a) shows the zoomed view of higher Chern number assisted Weyl nodes at $\Gamma$ and $R$ point in the BZ.}}
\label{fig5}
\end{figure}

\begin{figure*}[t]
\includegraphics[width=0.9\linewidth]{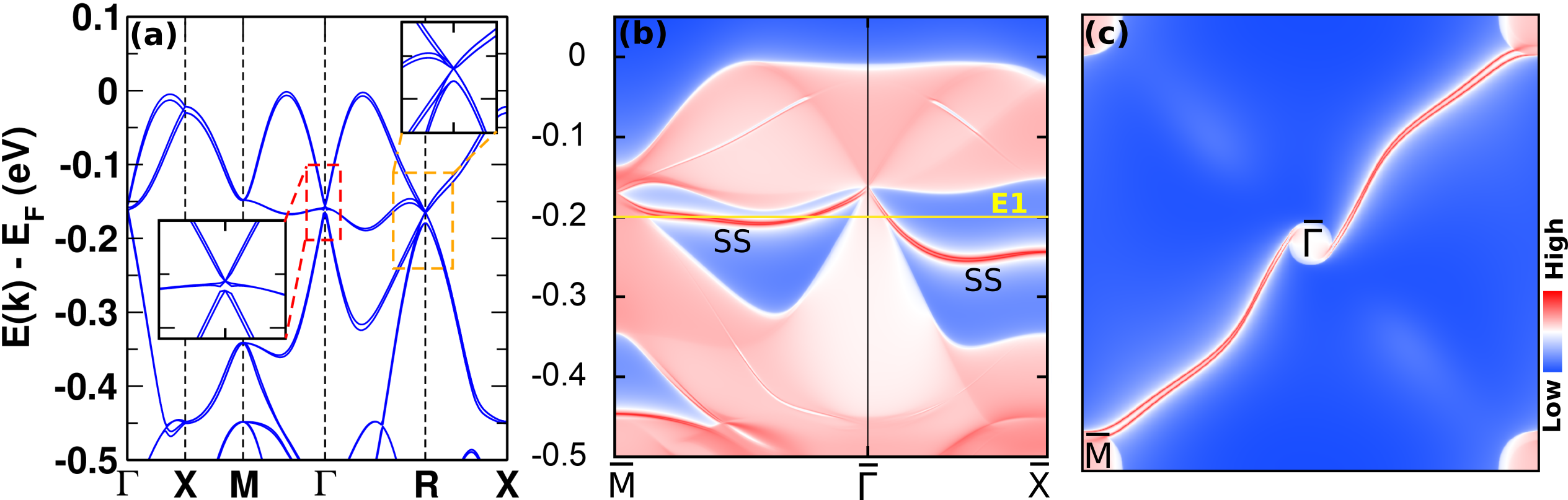}
\caption{(Color online) \textcolor{black}{For KMgBO$_3$ with SOI: (a) Bulk band structure, (b) Surface spectrum at side surface (001). Surface states are marked by SS. (c) Fermi arc contour at energy E1=E$_F - 0.2$ eV, shown by the horizontal yellow line in (b). As mentioned in the text, SOI-induced splitting is close to imperceptible (compare with Fig. S5 of SI \cite{supp})}}
\label{fig6}
\end{figure*}

{\par} \SPEDITOKAY{Next, we include the effect of SOI into our calculations}{The effect of SOI is expected to be relevant for CoGe and BiSbPt}, and the corresponding spinfull results are shown in  the rest of 
Fig.~\ref{fig3} and Fig.~\ref{fig5}(a). At $\Gamma$, we get at most a four-fold degeneracy as dictated by symmetry arguments. One such four-fold degeneracy is highlighted in Fig.~\ref{fig3}(b). Whereas at $R$-point, six-fold degeneracy is also allowed by symmetries as highlighted in Fig.~\ref{fig3}(b). Figure~\ref{fig3}(c) illustrates the Fermi surface (FS) map with SOI. At $\Gamma$ point, two concentric spherical shape FSs are found, which arise from the four-fold spin-3/2 excitations. The bands in the inner(outer) sphere possess Chern number -1(-3). At $R$, FS corresponds to four electron-like bands from double spin-1 excitations with C=$\pm2$. Along $\Gamma$-R and at M point in the BZ, tiny Fermi pockets are observed. 
We further show the Berry curvature ($\vec{\Omega}$) on k$_z$=0 and k$_x$=k$_y$ plane in Fig.~\ref{fig3}(d,e) to highlight that it flows between $R$ and $\Gamma$ points in agreement with the sign of the topological charges. Notably, under ambient conditions, CoGe crystallizes in the SG $C2/m$,\cite{CoGeSPG92,CoGefootnote} where none of the above band topology is observed in our calculations.\cite{supp}

\par {\it Surface Excitations}:\
Figure~\ref{fig4} shows the surface state results for these unconventional fermions. Fermi arcs (FAs) on the surface, if present, are generally expected to connect topological nodes of opposite chirality. We studied the (001) surface in which R and $\Gamma$ points fall at different locations (as shown in Fig.~\ref{fig1}(c)) as opposed to (111) surface to allow for large FAs. 
In presence of SOI \SPEDITOKAY{doubles}{and consequent doubling of} the Chern number ($|C|=4$) at R and $\Gamma$ point, \SPEDITOKAY{the essential new features involve}{there are} two pairs of FA states that emerge from the bulk projected states at $\bar{\Gamma}$ and  $\bar{M}$ point, as seen \AAA{from the surface spectrum shown} in Fig.~\ref{fig4}(a). \CMEDITOKAY{}{FA spectral weights are shown in Fig.~\ref{fig4}(b,c) at two different energy cuts.} Fig.~\ref{fig4}(b) also reveals the spin-momentum-locked spin texture of the FAs in the presence of SOI. Without SOI, two doubly-spin-degenerate FAs are present (see SI\cite{supp} for more details). SOI lifts the spin-degeneracy everywhere except at time-reversal invariant momenta, and thus two pairs of FAs appear with anti-parallel spin polarization. Such spin polarized textures may offer applications in spintronics.\cite{spintronics2019,SParkin2010}

\subsection{Beyond Binary Compounds}
\textcolor{black}{ We now report the simulated results of prototype ternary and quaternary systems -- BiSbPt and KMgBO$_3$ -- that belong to the same SG as the binary CoGe. Figure ~\ref{fig5}(a) display the bulk band structure of BiSbPt in presence of SOI. As expected, it shows various higher-fold fermions in concurrence with our general symmetry arguments. Figure~\ref{fig5}(b) shows the \SPEDITOKAY{topological signatures}{\textcolor{black}{FAs}} on (001) surface originating from these four-fold and six-fold Weyl nodes in bulk. BiSbPt hosts four pairs of surface states near E$_F$ (shown as SS1 \& SS2 in Fig. \ref{fig5}b). SS1 states emerge from the spin-3/2 node just above the E$_F$, while the SS2 states emerge from the spin-3/2 node at around -0.26 eV below the        E$_F$ (see Fig. \ref{fig5}a).}
\textcolor{black}{In contrast to Ref. \onlinecite{RhSi2017}, where only bulk properties of few ternary compounds are shown, the multi-fold degenerate Weyl nodes in our predicted BiSbPt compound lie almost at the Fermi level and the extra trivial Fermi pockets are nearly absent. This, in turn, yields clean surface states near E$_F$ (see Fig.~\ref{fig5}(b)), however there are comparatively more spectral weights arising from the bulk than CoGe.}

\textcolor{black}{Remarkably, we found that the quaternary compound KMgBO$_3$ from the orthoborate family shows the cleanest FAs when compared to all the systems we studied as well as CoSi from the previous report. KMgBO$_3$ has already been synthesized using solid-state reaction techniques without requiring high pressures.\cite{KMgBO3-2010} It is expected to have weak SOI because of its light constituent elements. Figure ~\ref{fig6} shows the bulk band structure and surface spectra with SOI. As clearly visible, there is a pair of almost degenerate large FAs running from $\bar{\Gamma}$ to $\bar{M}$ with almost no mixing from the bulk states, thus making KMgBO$_3$ an exciting candidate for future experimental studies. The weak-SOI nature of KMgBO$_3$ is corroborated by the negligible effect of SOI on both the bulk and surface electronic structure. The maximal SOI-induced splitting is less than 0.01 eV, and the bulk and surface electronic structure is essentially a ``doubled copy" of the corresponding spinless band structure (see Fig. S5 of SI\cite{supp}). We note here that SOI effects are already very small for CoSi\cite{CoSiARPES2019,CoSi2019,CoSi2019ARPES} and this should carry over for KMgBO$_3$ as well. The degeneracies in the bulk are again in accordance with earlier symmetry considerations, and can essentially be understood using the spinless arguments. Moreover, we also found several other experimentally synthesized quaternaries (Ag$_4$Te(NO$_3$)$_2$, Ag$_4$Te(ClO$_3$)$_2$ and Ag$_4$TeSO$_4$) with SG 198 and they again show three, four and six-fold degenerate Weyl nodes at $\Gamma$ and R point in the BZ. The multi-Weyl nodes in these quaternary systems also lie quite close to E$_F$ (see Fig. S6 of SI).\cite{supp}}

\section{Conclusion}
It is  important to note that the Weyl nodes that appear in systems such as WTe$_2$,\citep{Alexey2015} MoTe$_2$,\citep{MoTe2} LiAlGe,\citep{LiAlGe} TaAs(P)\citep{TaAsNbAs}, NbAs(P)\citep{TaPNbP} and so on are accidental band crossings with the FAs relatively smaller in size. In contrast, the $\Gamma$ and $R$ point band-crossings in CoGe, BiSbPt and KMgBO$_3$ (all belonging to SG 198) are robustly protected by the crystal space group symmetries. Also the FAs on the (001) surface \CMEDITOKAY{CoGe}{} are much larger since the nodes are maximally separated in BZ. Another promising feature of these systems -- especially CoGe and KMgBO$_3$ --  is the ``clean" nature of FAs because of the near absence of spectral weights from bulk states at $E_F$ as evident from Figs.~\ref{fig4},\ \ref{fig5} and \ref{fig6}. This makes them relatively superior than many other reported binary alloys (of SG 198), such as GaPt,\citep{GaPt} GaPd,\citep{GaPd} AlPd,\citep{RhSi2017} AlPt,\citep{RhSi2017,AlPt2018} RhGe,\citep{RhSi2017} AuBe,\cite{AuBe} $M$Si ($M$=Fe, Mn, Ru, Re)\cite{TZHANG2018} which suffer from large spectral weight contributions of extra bulk band crossings across E$_F$. Very recently, experiments\cite{CoSiARPES2019,CoSi2019,CoSi2019ARPES} have borne out these advantages for the related compound CoSi,\cite{SCZHANG2017} which makes the case for experiments on CoGe and KMgBO$_3$ attractive since they have already been successfully synthesized.\cite{CoGeExpt,KMgBO3-2010}

In summary, we predict an ideal higher Chern-number topological semimetal in CoGe in agreement with previous bulk studies on binary systems with SG 198. We showed giant FA states in this system without much contamination from the bulk states. Furthermore, we have identified the existence of four and six fold degenerate Weyl nodes and their novel surface signatures in a ternary BiSbPt, and a quaternary compound KMgBO$_3$. These unconventional multi-Weyl nodes lie close to the Fermi level which make these {\it beyond binary systems} experimentally quite promising as well. KMgBO$_3$ additionally has exceptionally clean, giant FA states compared to all other systems as has been emphasized before. 
At a theoretical level, we gave new, alternate Kramer's theorem-like arguments based on the inter-relationships between two non-symmorphic screws and three-fold rotations of SG 198 to explain the four-fold degeneracies at $R$ point for the spinless case (only possibility) and at $\Gamma$ point for the spinfull case. Thus, they were expectedly seen in all the non-binary and binary systems with SG 198 studied by us. 
The energy offset observed between the multi-Weyl nodes at $\Gamma$ and R point makes these systems suitable for observing quantized circular photogalvanic effect with possibilities for technological applications.\cite{RhSi2017,photogalvanic} All these features of CoGe, BiSbPt and KMgBO$_3$ serve as strong motivation for future experimental investigations to study these candidate chiral semimetals with topological charges larger than C=$\pm$1.

\section*{Acknowledgement}
CKB acknowledges IIT Bombay for financial support in the form of teaching assistantship. CM acknowledges MHRD-India for financial support. AA acknowledges DST-SERB (Grant No. CRG/2019/002050) for funding to support this research. SP acknowledges financial support from IRCC, IIT Bombay (17IRCCSG011) and SERB, DST, India (SRG/2019/001419).

\appendix*
\section{Derivations of Band Degeneracies and Further details on Symmetry Arguments}

This appendix contains auxiliary elaborations on the symmetry arguments presented concisely in the main text. Section \ref{sec:some_preliminaries} sets up the preliminaries of symmetry operations. Section \ref{sec:degen_spinless} is devoted to the band degeneracies at $\Gamma$ and $R$ point for spinless case. Section \ref{sec:degen_spinfull} is devoted to the analysis of degeneracy for spinfull case. Section \ref{sec:line_degen} explains the twofold line degeneracies along $R$-$X$ and $M$-$X$ high symmetry directions in the BZ.

\subsection{Some Preliminaries}
\label{sec:some_preliminaries}
Following usual conventions, we will specify any crystal symmetry operation by 
a point group operation $\mathcal{O}$ followed by a translation, $\vec{t}$. For pure point group operations,
$\vec{t}=\left(0,0,0\right)$. The 
rules of combining two crystal symmetry operations is:
\begin{align*}
\textcolor{blue}{\{\mathcal{O}_1|\vec{t_1}\} \{\mathcal{O}_2|\vec{t_2}\}} \textcolor{blue}{=}\;& \textcolor{blue}{\{\mathcal{O}_{1}\mathcal{O}_{2}|\mathcal{O}_{1}\vec{t_2}+\vec{t_1}\}} \\
\textcolor{magenta}{\{\mathcal{O}|\vec{t}\}^{-1}} \textcolor{magenta}{=}\;& \textcolor{magenta}{\{\mathcal{O}^{-1}|-\mathcal{O}^{-1}\vec{t}\}} \\
\end{align*}

Pure translations are indicated by \textcolor{ForestGreen}{$\{\mathbb{I}|\vec{t}\} = e^{-i\vec{k}.\vec{t}}$}, 
where $\mathbb{I}$ is an identity operation, and $\vec{k}$ and $\vec{t}$ are reciprocal wave vector 
and translation vector respectively. We use $\mathcal{R}$ to signify a $2\pi$ rotation,
which equals $\mathbb{I}$ and $-\mathbb{I}$ for spinless and spinfull cases respectively.

The two-fold ($C_2$ ) and three-fold ($C_3$ ) rotation operators transform lattice co-ordinates as follows:
\begin{eqnarray*}
\textcolor{violet}{C_{2x}\left(x,y,z \right )\longrightarrow \left(x,-y,-z \right ) } \\
\textcolor{violet}{C_{2y}\left(x,y,z \right )\longrightarrow \left(-x,y,-z \right )} \\
\textcolor{violet}{C_{2z}\left(x,y,z \right )\longrightarrow \left(-x,-y,z \right )} \\
\textcolor{violet}{C_{3,111}\left(x, y, z \right )\longrightarrow  \left(z, x, y \right )} \\
\textcolor{violet}{C^{-1}_{3,111}\left(x, y, z \right )\longrightarrow  \left(y, z, x \right )} 
\end{eqnarray*}
The matrix representations of these rotation operators are thus as follows:
\begin{eqnarray*}
C_{2x}&=&\begin{pmatrix} 
1 & 0 & 0 \\
0 & -1 & 0  \\
0 & 0 & -1 
\end{pmatrix};
C_{2y}=\begin{pmatrix} 
-1 & 0 & 0 \\
0 & 1 & 0  \\
0 & 0 & -1 
\end{pmatrix}\\
C_{2z}&=&\begin{pmatrix} 
-1 & 0 & 0 \\
0 & -1 & 0  \\
0 & 0 & 1 
\end{pmatrix};
C_{3,111}= \begin{pmatrix} 
0 & 0 & 1 \\
1 & 0 & 0  \\
0 & 1 & 0 
\end{pmatrix}
\end{eqnarray*}
and we can use them to multiply rotation operators 
(\textcolor{red}{ \{$\mathcal{O}_1\mathcal{O}_2\mathcal{O}_3...$\}}) to obtain the net point group operation. Sum of two translation vectors follows the usual rule:
\begin{eqnarray*}
\textcolor{ginger}{\left(x_1,y_1,z_1 \right )+\left(x_2,y_2,z_2 \right )\longrightarrow \left(x_1+x_2,y_1+y_2,z_1+z_2\right ) } \\
\end{eqnarray*}
Furthermore, the color scheme set up above will be used
in the remaining text when needed to allow for easy parsing of the various algebraic manipulations. In some algebraic manipulation, any expression with a given color in any line is replaced in the following line by the right hand side of the corresponding colored formula above.

\subsection{Spinless Case}
\label{sec:degen_spinless}

\subsubsection{$\Gamma$ point}
\label{subsec:spinlessG}
The little group at $\Gamma$ point has $S_{2z}=\{C_{2z}|\frac{1}{2},0,\frac{1}{2}\}$, $S_{2y}=\{C_{2y}|0,\frac{1}{2},\frac{1}{2}\}$ \& $S_3=\{C_{3,111}|0,0,0\}$ as the symmetry generators.\cite{CracknellBook} 
These generators satisfy the following relations:
\begin{eqnarray}\label{GammaS2z2}
S^{2}_{2z} &=& \textcolor{blue}{\{C_{2z}|\frac{1}{2},0,\frac{1}{2}\}\{C_{2z}|\frac{1}{2},0,\frac{1}{2}\}} \nonumber \\
          &=& \{C^{2}_{2z}|\textcolor{violet}{C_{2z}\left(\frac{1}{2},0,\frac{1}{2} \right )}+\left(\frac{1}{2},0,\frac{1}{2} \right )\}  \nonumber \\
          &=& \{C^{2}_{2z}|\textcolor{ginger}{\left(\bar{\frac{1}{2}},0,\frac{1}{2} \right )+\left(\frac{1}{2},0,\frac{1}{2} \right )}\} \nonumber \\
	  &=&\{\textcolor{red}{C^{2}_{2z}}|0,0,1\} \nonumber \\
	  &=&\{\mathcal{R}|0,0,1\} \nonumber \\
	  &=&\textcolor{ForestGreen}{\{\mathbb{I}|0,0,1\}} \nonumber \\
          &=& 1
\end{eqnarray}
From now onwards, we will skip the derivations of the various relations satisfied by the crystal symmetries,
and only focus on the details of the symmetry-protection of the degeneracies.
All derivations of crystal symmetry relations 
are compiled in Sec. VII of SI.\cite{supp}
Similar to Eq. \ref{GammaS2z2}, we also get
\begin{eqnarray}\label{GammaS2y2}
	S^{2}_{2y} = 1 \\
	\label{GammaS33}
S^{3}_{3} = 1
\end{eqnarray}

The two-fold screws and three-fold rotation $C_{3,111}$ satisfy the following relations:
\begin{subequations}
\label{GammaScrews}
\begin{align}
[ S_{2z},S_{2y}] &=  0 \label{Gammacommutator}\\
S_{2z}S_{3} &=  S_{3}S_{2y} \label{GammaS2zS3} \\
S_{3}S_{2z}S_{2y} &=  S_{2y}S_{3} \label{GammaS3S2zS2y}
\end{align}
\end{subequations}

Since $S_{2z}$ and $S_{2y}$ commute, let $|\psi\rangle$ be a 
simultaneous eigenstate of both $S_{2z}$ and $S_{2y}$ (and also the Hamiltonian since these 
are the symmetries
of the Hamiltonian, i.e. commute with the Hamiltonian by definition).\\ Let
\begin{eqnarray}
S_{2z}|\psi\rangle &=& \lambda_1|\psi\rangle \nonumber \\
S_{2y}|\psi\rangle &=& \lambda_2|\psi\rangle
\end{eqnarray}
with $\lambda_1=\pm1$, $\lambda_2=\pm1$ due to Eqs. \ref{GammaS2z2} and \ref{GammaS2y2}.

Using above relations between $S_3$, $S_{2z}$ and $S_{2y}$, we can arrive at
\begin{eqnarray}\label{10}
S_{2z}S_{3}|\psi\rangle &=& S_{3}S_{2y}|\psi\rangle =  \lambda_2 S_{3}|\psi\rangle \\
S_{2y}S_{3}|\psi\rangle &=& S_{3}S_{2z}S_{2y}|\psi\rangle = \lambda_1\lambda_2 S_{3}|\psi\rangle \nonumber \\
S_{2z}S^2_{3}|\psi\rangle &=& S_{3}S_{2y}S_{3}|\psi\rangle = S^2_{3}S_{2z}S_{2y}|\psi\rangle = \lambda_1\lambda_2 S^2_{3}|\psi\rangle \nonumber \\
S_{2y}S^2_{3}|\psi\rangle &=& S_{3}S_{2z}S_{2y}S_{3} |\psi\rangle = S^2_{3}S_{2z}|\psi\rangle = \lambda_1 S^{2}_{3}|\psi\rangle \nonumber
\end{eqnarray}
The set of equations Eqs. \eqref{10} show $S_3$ generates two new distinct eigenstates 
$S_{3}|\psi\rangle$ and $S^{2}_{3}|\psi\rangle$ of $S_{2z}$ and $S_{2y}$ 
provided either $\lambda_1 \neq 1$ or $\lambda_2 \neq 1$. In other words, both screws are non-trivial. 
These three states will be degenerate since
$S_3$ commutes with the Hamiltonian.
Thus, these three states ($|\psi\rangle$, $|S_{3}\psi\rangle$, $|S^{2}_{3}\psi\rangle$ ) 
together form a three-fold degeneracy at $\Gamma$ point. 
The above is a recapitulation of the arguments in Sec. C in Ref. \onlinecite{Bradlyn2016}'s
supplementary.
The $\lambda_1=\lambda_2=1$
may correspond to a case where both screws are trivial which does not protect any degeneracy, or a case where only one of the screws is trivial which protects only a two-fold degeneracy.

\subsubsection{$R$ point}
\label{subsec:spinlessR}
The generators at R point are 
$S_{2x}=\{ C_{2x}|\frac{1}{2},\frac{3}{2},0\}$, $S_{2y}=\{C_{2y}|0,\frac{3}{2},\frac{1}{2}\}$ 
and $S_{3}=\{C^{-1}_{3,111}|0,1,0\}$.\cite{CracknellBook}
They satisfy the following:
\begin{subequations}
\label{RScrews}
\begin{align}
S^{2}_{2x} = {} & -1 \label{RS2x2} \\
S^{2}_{2y} = {} &-1 \label{RS2y2} \\
S_{2x}S_{2y} = {} & -S_{2y} S_{2x} \label{RS2xS2y} \\
S_{2x}S_{3} = {} & S_{3} S_{2y} \label{RS2xS3} \\
S_{3}S_{2x}S_{2y} = {} & S_{2y}S_{3} \label{RS3S2xS2y}
\end{align}
\end{subequations}

The eigenvalues under the two-fold screws ($S_{2x},S_{2y}$) will be unit-modulus and pure imaginary
due to Eqs. \ref{RS2x2} and \ref{RS2y2}. Let $|\psi\rangle$ be an eigenstate of $S_{2y}$ 
with eigenvalue $i$ without loss of generality, i.e. $S_{2y} |\psi\rangle = i |\psi\rangle$.
Then, 
Eq. \ref{RS2xS2y} implies that $|S_{2x} \psi\rangle \equiv S_{2x} |\psi\rangle$ 
will be an eigenstate of $S_{2y}$ with eigenvalue $-i$ because
\begin{equation*}
\begin{aligned}
S_{2y}|S_{2x} \psi\rangle =& S_{2y} S_{2x} |\psi\rangle = - S_{2x} S_{2y} |\psi\rangle = - i |S_{2x} \psi\rangle
\end{aligned}
\end{equation*} 
Since $|\psi\rangle$ and $|S_{2x} \psi\rangle$ have different eigenvalues under $S_{2y}$, they are orthogonal.
Eq. \ref{RS2xS3} now implies that $|S_3 \psi\rangle \equiv S_3 |\psi\rangle$ will be an 
eigenstate of $S_{2x}$ with eigenvalue $i$ because
\begin{equation*}
\begin{aligned}
S_{2x} |S_3 \psi\rangle = {} & S_{2x} S_3 |\psi\rangle = S_3 S_{2y} |\psi\rangle = S_3 i |\psi\rangle 
	= i |S_3 \psi\rangle
\end{aligned}
\end{equation*}
Eq. \ref{RS2xS2y} will again imply that $|S_{2y} S_3 \psi\rangle \equiv S_{2y} S_3 |\psi\rangle$ 
will be an eigenstate of $S_{2x}$ with eigenvalue $-i$ because 
\begin{equation*}
\begin{aligned}
S_{2x} |S_{2y} S_3 \psi\rangle = {} &  S_{2x} S_{2y} |S_3 \psi\rangle = - S_{2y} S_{2x} |S_3 \psi\rangle 
	= - i |S_{2y} S_3 \psi\rangle
\end{aligned}
\end{equation*}
Since $|S_3 \psi\rangle$ and $|S_{2y} S_3 \psi\rangle$ have different eigenvalues under $S_{2x}$, they are orthogonal.
\cite{extra_footnote}

By requiring that $S_3$ acts non-trivially on the eigenstates of $S_{2y}$ and takes 
out of the subspace formed by them, we can ensure mutual orthogonality between eigenstates of $S_{2y}$ and $S_{2x}$. 
Minimally, $\langle\psi | S_3 \psi\rangle =0$ guarantees all other mutual orthogonalities as follows:
{\par} Case of $|S_3 \psi\rangle $ and $|S_{2x} \psi\rangle$: 
\begin{equation*}
\begin{aligned}
\langle S_{2x} \psi | S_3 \psi\rangle = & \langle\psi | S_{2x}^{-1} S_3 | \psi\rangle = \langle\psi | \left(-S_{2x} \right ) S_3 | \psi\rangle \\
= & - \langle\psi | S_{2x} S_3 | \psi\rangle = \langle\psi | S_3 S_{2y} | \psi\rangle \\
=& i \langle\psi | S_3 | \psi\rangle = 0.
\end{aligned}
\end{equation*}
{\par} Case of $|S_{2y} S_3 \psi\rangle$ and $|\psi\rangle$:
\begin{equation*}
\begin{aligned}
\langle\psi | S_{2y} S_3 \psi\rangle =& \langle\psi | S_{2y} S_3 | \psi\rangle 
= - \langle\psi | S_{2y}^{-1} S_3 | \psi\rangle \\
=& - \left(-i \right ) \langle\psi | S_3 | \psi\rangle = 0
\end{aligned}
\end{equation*}
{\par} Case of $|S_{2y} S_3 \psi\rangle$ and $|S_{2x} \psi\rangle$:
\begin{equation*}
\begin{aligned}
\langle S_{2x} \psi | S_{2y} S_3 \psi\rangle =& \langle\psi | S_{2x}^{-1} S_{2y} S_3 | \psi\rangle \\
=& \langle\psi | \left(-S_{2x} \right ) S_{2y} S_3 | \psi\rangle \\
=& - \langle\psi | S_{2x} |S_{2y} S_3 \psi\rangle \\
=& i \langle\psi | S_{2y} S_3 \psi\rangle = 0
\end{aligned}
\end{equation*}
Therefore, $\left(|\psi\rangle, |S_{2x} \psi\rangle, |S_3 \psi\rangle, |S_{2y} S_3 \psi\rangle \right)$ 
are four mutually orthogonal states. Thus, we have a symmetry-protected four-fold 
degeneracy at $R$ point in the absence of spin-orbit coupling.

Since time-reversal squares to identity ($\mathcal{T}^2 = \mathbb{I}$) for spinless fermions, 
it does not generate any new eigenstates. In fact, it relates the eigenstates of the two screws as follows:
\begin{align*}
S_{2y} |\mathcal{T} \psi\rangle =\;& S_{2y} \mathcal{T} |\psi\rangle = \mathcal{T} S_{2y} |\psi\rangle 
= \mathcal{T} i |\psi\rangle = -i \mathcal{T} |\psi\rangle \\
\Rightarrow S_{2y} \;& |\mathcal{T} \psi\rangle = -i |\mathcal{T} \psi\rangle
\end{align*}
where we have used the facts that $\mathcal{T}$ commutes with the screws, 
and $\mathcal{T}^\dagger i \mathcal{T} = -i$ (anti-linear property).
Thus, we can identify $|\mathcal{T} \psi\rangle$ with $|S_{2x} \psi\rangle$ having same eigenvalue $-i$
under $S_{2y}$.
By a very similar argument, the pairs $\{|\mathcal{T} S_{2x} \psi\rangle, |\psi\rangle\}$,
$\{|\mathcal{T} S_{3} \psi\rangle, |S_{2y} S_3 \psi\rangle\}$ and
$\{|\mathcal{T} S_{2y} S_3 \psi\rangle, |S_3|\psi\rangle\}$
can be identified. 

To demonstrate the four-fold degeneracy at $R$-point for spinless case, we have simulated few more binary and ternary systems belonging to the space group $198$. The bulk band structure for these compounds are shown in Fig. S3 of supplement.\cite{supp} Similar to CoGe, the electronic structure in all these binary and ternary systems shows the four-fold degeneracy at $R$ point irrespective of their location with respect to Fermi level. Thus, the four-fold degeneracy at $R$ point for the spinless case is independent of both the chemical elements at the lattice sites and number of atoms in the cell. Rather, the degeneracy at R point is strictly determined by the crystal space group symmetry.

\subsection{Spinfull Case}
\label{sec:degen_spinfull}

The generators at $\Gamma$ point are $S_{2z}=\{C_{2z}|\frac{1}{2},0,\frac{1}{2}\}$, $S_{2y}=\{C_{2y}|0,\frac{1}{2},\frac{1}{2}\}$ \& $S_3=\{C_{3,111}|0,0,0\}$.\cite{CracknellBook} They satisfying the following relations for spinfull fermions:

\begin{subequations}
\label{GammaScrewsSOC1}
\begin{align}
S^{2}_{2z} =  &  -1 \label{GammaS2z2SOC} \\
S^{2}_{2y} = &  -1 \label{GammaS2y2SOC} \\
S^{3}_{3} 
	= & - 1 \label{GammaS33SOC}
\end{align}
\end{subequations}

The difference with respect to the corresponding spinless $\Gamma$ point symmetry relations 
is due to the different action of $\mathcal{R}$ in these two cases.

Also, we have
\begin{subequations}
\label{GammaScrewsSOC2}
\begin{align}
S_{2z}S_{2y} = {} &   -S_{2y}S_{2z} \label{GammacommutatorSOC}\\
S_{2z}S_{3} = {} &  S_{3}S_{2y} \label{GammaS2zS3SOC} \\
S_{3}S_{2z}S_{2y} = {} & S_{2y}S_{3} \label{GammaS3S2zS2ySOC}
\end{align}
\end{subequations}
Therefore, we can use the very same arguments as in Sec. \ref{subsec:spinlessR} to generate a four-fold degeneracy.

Since $\mathcal{T}^2 = -1$ for the spinfull case, so it is possible that time reversal may generate further new states.
In other words, the question is whether the time-reversed partners of the above four-fold states \{$|\psi\rangle, |S_{2z} \psi\rangle, |S_3 \psi\rangle, |S_{2y} S_3 \psi\rangle$\} are distinctly new states or not? 
As mentioned in the main text, they are actually not new states because mutual orthogonalities are not ensured.
This is due to the imaginary eigenvalues under screws. 

From $S_{2z}^2 = S_{2y}^2 = \mathcal{T}^2 = -1$, we have $S_{2z}^{-1} = - S_{2z}$, $S_{2y}^{-1} = - S_{2y}$ and $\mathcal{T}^{-1} = -\mathcal{T}$. Also $\mathcal{T}$ commutes with the screws. Firstly, these mutual overlaps have
to be real. E.g.
\begin{equation*}
\begin{aligned}
	& \langle S_{2z} \psi | \mathcal{T} \psi\rangle  \nonumber \\
	= & \langle\psi | S_{2z}^{-1} \mathcal{T} | \psi\rangle \nonumber
=  - \langle\psi | S_{2z} \mathcal{T} | \psi\rangle \nonumber
= - \langle\psi | \mathcal{T} S_{2z} | \psi\rangle \nonumber \\
	= & \langle\psi | \mathcal{T}^{-1} S_{2z} | \psi\rangle \nonumber 
= \langle\mathcal{T} \psi | S_{2z} \psi\rangle \nonumber 
= \langle S_{2z} \psi | \mathcal{T} \psi\rangle^{*}
\end{aligned}
\end{equation*}

Secondly, the eigenvalue of $|\mathcal{T} \psi\rangle$ under $S_{2y}$ is same as $|S_{2z} \psi\rangle$, and similarly the eigenvalue of $|\mathcal{T} S_{2z} \psi\rangle$ under $S_{2y}$ is same as $|\psi\rangle$ as follows: 
Let, $S_{2y} |\psi\rangle = i |\psi\rangle$
Therefore, \[ S_{2y} |S_{2z} \psi\rangle = -i |S_{2z} \psi\rangle\] by following the same argument as in Sec. \ref{subsec:spinlessR}. Now, 
\begin{equation*}
\begin{aligned}
S_{2y} |\mathcal{T} \psi\rangle =& S_{2y} \mathcal{T} |\psi\rangle = \mathcal{T} S_{2y} |\psi\rangle = \mathcal{T} i |\psi\rangle \\
=& -i \mathcal{T}  |\psi\rangle = -i |\mathcal{T} \psi\rangle.
\end{aligned}
\end{equation*}
Thus, both $|\mathcal{T} \psi\rangle$ and $|S_{2z} \psi\rangle$ have the same eigenvalues under $S_{2y}$, and we can not conclude anything about this mutual orthogonality. The same lack of mutual orthogonality 
will be the case for the other pairs 
$\{|\psi\rangle,|\mathcal{T} S_{2z} \psi\rangle\}$, $\{|S_3 \psi\rangle,|\mathcal{T} S_{2y} S_3 \psi\rangle\}$
and $\{| S_{2y} S_3 \psi\rangle, |\mathcal{T}S_3 \psi\rangle\}$. 
Thus, we can at most get a four-fold degeneracy 
($\{ |\psi\rangle, |S_{2z} \psi\rangle, |S_3 \psi\rangle, |S_{2y} S_3 \psi\rangle\}$)
at $\Gamma$ point for spinfull fermions in SG 198.

We also note here that for $R$ point, since now the screws commute
and square to 1, the eigenvalues are unit modulus and purely real. 
We can get a three-fold degeneracy ($\{|\psi\rangle, |S_3 \psi\rangle, |S^2_3 \psi\rangle\}$) 
by following the same arguments as in Sec. \ref{subsec:spinlessG}.
Furthermore due to eigenvalues being real, the above mutual orthogonalities under time reversal
are ensured,
and we have three distinctly new time-reversed partners 
($\{|\mathcal{T}\psi\rangle, |\mathcal{T}S_3 \psi\rangle, |\mathcal{T}S^2_3 \psi\rangle\}$).
This can give a symmetry-protected six-fold degeneracy at $R$ spinfull case as 
discussed in Ref. \onlinecite{Bradlyn2016}.

\subsection{Two fold degeneracies along R-X and M-X direction}
\label{sec:line_degen}

\subsubsection{Spinless Case}
\label{subsec:line_spinless}
The screw rotation along $x$-axis is $S_{2x} = \{C_{2x}|\frac{1}{2},\frac{1}{2},0\}$.
We can define an anti-unitary operator $\Theta_{2x}=\mathcal{T}S_{2x}$.
$\mathcal{T}$ squares to $+1$ for spinless case, and commutes with the (unitary) screw.
Thus we have
\begin{eqnarray}\label{twofold}
\Theta^{2}_{2x} &=&  \mathcal{T} S_{2x} \mathcal{T} S_{2x} \nonumber \\
&=&  \mathcal{T}^2 S^2_{2x} \nonumber \\
&=& \mathcal{T}^2 \textcolor{blue}{ \{C_{2x}|\frac{1}{2},\frac{1}{2},0\} \{C_{2x}|\frac{1}{2},\frac{1}{2},0\}} \nonumber \\
&=& \mathcal{T}^2 \{ C^2_{2x}|\textcolor{violet}{ C_{2x}\left(\frac{1}{2},\frac{1}{2},0 \right )}+\left(\frac{1}{2},\frac{1}{2},0 \right )\} \nonumber \\
&=& \mathcal{T}^2 \{ C^2_{2x}|\textcolor{ginger}{ \left(\frac{1}{2},\bar{\frac{1}{2}},0 \right )+\left(\frac{1}{2},\frac{1}{2},0 \right )}\} \nonumber \\
	&=& \mathcal{T}^2 \{\textcolor{red}{C^2_{2x}}|1,0,0\} \nonumber \\
	&=& \textcolor{red}{\mathcal{T}^2} \{ \mathcal{R}|1,0,0\} \nonumber \\
&=& \{ \mathcal{R}|1,0,0\} \nonumber \\
&=& \textcolor{ForestGreen}{\{ \mathbb{I}|1,0,0\}} \nonumber \\
&=& e^{-ik_x}
\end{eqnarray}

Therefore, on the $k_x = \pi$ plane, $\Theta^{2}_{2x}=-1$. Thus, by Kramer's argument, 
if $|\psi\rangle$ is an eigenstate of $S_{2x}$, then $|\Theta_{2x} \psi\rangle$ 
is a like a time-reversed partner for $k_x = \pi$. 
Hence, $\Theta_{2x}$ gives a Kramer's like double degeneracy on the $k_x = \pi$ and symmetry-related planes. 
This in turn implies that the bands along R-X and M-X directions in the Brillouin zone 
are two-fold degenerate by the combination of time-reversal and screw symmetry
as seen in Fig. 2(a) of the main text.

\subsubsection{Spinfull Case}
\label{subsec:line_spinfull}

For spinfull case, 
$\mathcal{R}=-\mathbb{I}$ and $\mathcal{T}^2=-\mathbb{I}$. 
Therefore, similar to Eq. \eqref{twofold}, it follows that

\begin{eqnarray*} 
\Theta^{2}_{2x} &=&  \mathcal{T} S_{2x} \mathcal{T} S_{2x}  \nonumber \\
&=&  \mathcal{T}^2 S^2_{2x} \nonumber \\
	&=& \mathcal{T}^2 \{ \textcolor{red}{C^2_{2x}}|1,0,0\} \nonumber \\
	&=& \textcolor{red}{\mathcal{T}^2} \{ \mathcal{R}|1,0,0\} \nonumber \\
&=& -\mathbb{I} \{ -\mathbb{I}|1,0,0\} \nonumber \\
&=& +e^{-ik_x}
\end{eqnarray*}
Thus, similar to the spinless case, $\Theta^{2}_{2x}=-1$ again and the bands are doubly degenerate on $k_x=\pi$ and
symmetry-related planes even in the spinfull case. 
These gives the double degeneracy of bands along R-X and M-X and symmetry-related directions
in the Brillouin zone as also observed in Fig. 2(b) of the main text.
We note here that this is again a Kramer's-like degeneracy ensured by
a combination of time reversal and screw symmetry on these planes, and
not the standard Kramer's degeneracy which can not be applied here since
inversion symmetry is absent. Screw symmetry is replacing the inversion symmetry
on these high-symmetry planes to again make the Kramer's argument operational
and give us a Kramer's-like two-fold degeneracy.




\begin{thebibliography}{113}
\bibitem{chern_footnote}
The topological charges or Chern numbers being integers can not change continuously.
\bibitem{s3_footnote} A trivial $S_3$, i.e. $S_3 |\psi\rangle = |\psi\rangle$, is not allowed, since two anticommuting operators can not share a simultaneous eigenstate with non-zero eigenvalue. $S_3$ can not exchange $|\psi\rangle$ and $S_{2x}|\psi\rangle$ as well for the same reason. $S_3$ can however in principle create some linear combination of $|\psi\rangle$ and $S_{2x}|\psi\rangle$, in which case this argument would give a two-fold degeneracy.
\bibitem{2pirot} This difference in screw commutation relations at $\Gamma$ and $R$ points is due to the different action of a $2\pi$ rotation for spinless and spinfull cases.
\bibitem{verbose} Verbose details are give in Appendix.
\bibitem{extra_footnote}
We also have $S_{2y} S_3  = S_3 S_{2x} S_{2y}$ at $R$ point, but this does not give any strong condition 
about the generation of new states.
\end{thebibliography}
\end{document}


\title{%
  Supplementary Information for `` Symmetry protection and giant Fermi arcs from mutifold fermions in binary, ternary and quaternary compounds"}

\author{Chanchal K. Barman}
\thanks{These two authors have contributed equally to this work}
\affiliation{Department of Physics, Indian Institute of Technology, Bombay, Powai, Mumbai 400076, India}

\author{Chiranjit Mondal}
\thanks{These two authors have contributed equally to this work}
\affiliation{Discipline of Metallurgy Engineering and Materials Science, IIT Indore, Simrol, Indore 453552, India}

\author{Sumiran Pujari}
\affiliation{Department of Physics, Indian Institute of Technology, Bombay, Powai, Mumbai 400076, India}

\author{Biswarup Pathak}
\email{biswarup@iiti.ac.in }
\affiliation{Discipline of Metallurgy Engineering and Materials Science, IIT Indore, Simrol, Indore 453552, India}
\affiliation{Discipline of Chemistry, School of Basic Sciences, IIT Indore, Simrol, Indore 453552, India}

\author{Aftab Alam}
\email{aftab@iitb.ac.in}
\affiliation{Department of Physics, Indian Institute of Technology, Bombay, Powai, Mumbai 400076, India}

\date{\today}
\maketitle
This supplement contains auxiliary
computational details, additional bulk and surface band structure calculations, and demonstrations of various algebraic relations between the crystalline symmetry.
 The computational details are given in Sec. \ref{computational}. 
Sec. \ref{sec:CoGe-nosoc} discusses about the surface excitations for CoGe in absence of spin-orbit interaction. In Sec. \ref{sec:C2/m}, we show the electronic structure for the case when CoGe crystallizes in space group $C2/m$ which has lower symmetry.  In Sec. \ref{subsec:examples}, we have shown band structures of several other binary and ternary alloys which 
belong to the same space group as CoGe. This provides additional corroboration on \emph{exclusive} four-fold degeneracies at $R$ point for any system that belongs to space group 198 in the absence of spin-orbit coupling. Sec. \ref{sec:BiSbPt} demonstrates about the Berry curvature, Fermi surface and Fermi arc for a ternary BiSbPt compound. In Sec. \ref{sec:quaternary}, we present band structures for few quaternary systems which are found to show three, four and six-fold degenerate Weyl nodes near the Fermi level as in the binaries and ternaries. Finally in Sec. \ref{sec:derivations} for reasons of pedagogical completeness, we have presented a compendium of the demonstrations of various algebraic relations between the crystal symmetry operations that are exploited to symmetry-protect the various degeneracies. 

\section{Computational Details}
\label{computational}
The {\it ab-initio} calculations were performed with projected augmented-wave basis\cite{PEBLOCHL1994} with an energy cut off of 500 eV. Total energy (force) was converged upto 10$^{-6}$ eV/cell (0.001 eV/\r{A}). A 12$\times$12$\times$12 $\Gamma$-centered k-mesh was used to perform the bulk Brillouin zone(BZ) integrations. The spin-orbit coupling (SOC) interaction was included self-consistently. Also as mentioned in the main text, Maximally Localized wannier functions (MLWF)\cite{MARZARI1997,SOUZA2001,VANDERBIT} were used to construct tight binding model Hamiltonian to closely reproduce the bulk band structure. The Chern numbers associated with the various multi-fold bands were calculated using the method of Wannier charge center (WCC) evolution implemented in wannier90 package.\cite{w90} The spectral weights in Fermi arcs and surface spectrum were calculated using iterative Green's function\cite{DHLEE_I,DHLEE_II,SANCHO1985} scheme implemented in WannierTools package.\cite{QuanSheng}

\section{Surface excitations for $\text{CoGe}$ in absence of spin-orbit interaction}
\label{sec:CoGe-nosoc}
\begin{figure}[ht!]
\centering
\includegraphics[width=\linewidth]{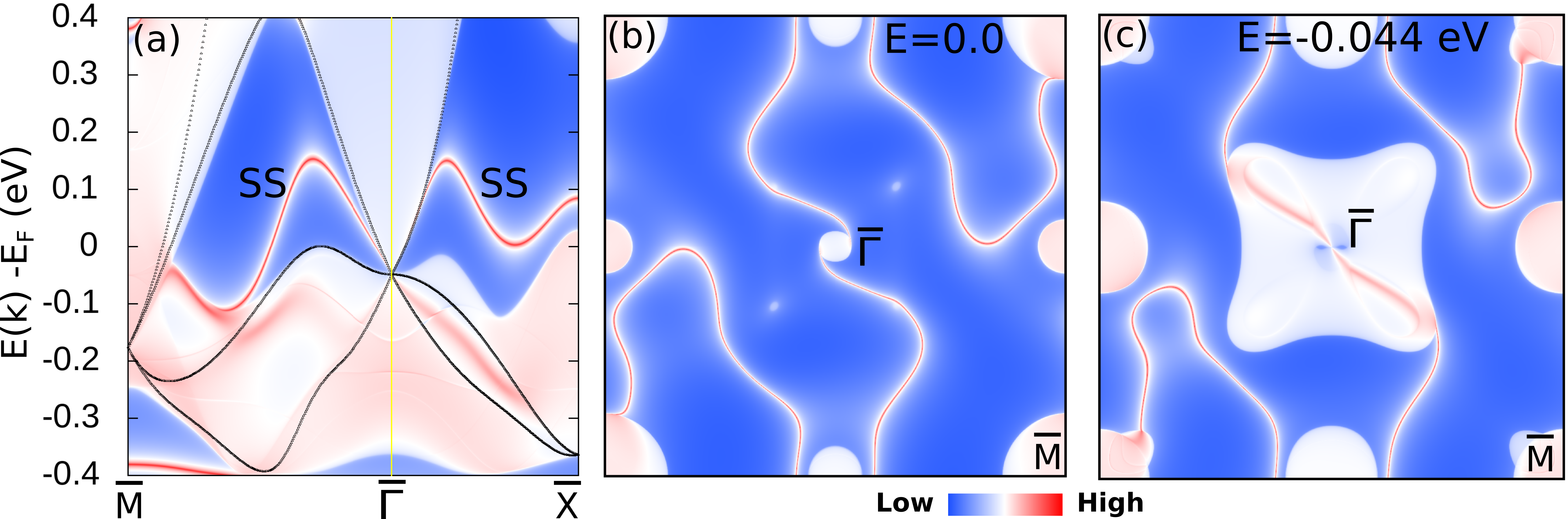}
\caption{(Color online) For CoGe without SOI; (a) surface spectrum at (001) side surface without SOI. Surface states are marked by SS. Superimposed bulk band structure along $R$-$\Gamma$-X are represented by black lines. (b,c) Fermi arc contour at energy window  E$_F$ (Fermi energy) and  E$_F -0.044$ eV (spin-1 Weyl node) without SOI.}
\label{figCoGe-nosoc}
\end{figure}

In Fig.~\ref{figCoGe-nosoc}, we show the surface signature of spin-1 and charge-2 Dirac node (see Fig. 2a in main manuscript) for CoGe in absence of spin-orbit interaction (SOI). We studied the surface state (SS) and Fermi arc (FA) topology on (001) surface in absence of SOI effect. Since the Chern number at $\Gamma $ is $2$, a pair of SS emerged from the $\bar{\Gamma}$ point as shown in Fig.~\ref{figCoGe-nosoc}(a). The FAs spectral weights in absence of SOI are shown in Fig.~\ref{figCoGe-nosoc}(b,c) at two different energy cuts (see caption). A pair of FAs runs between the $\bar{\Gamma}$ point and $\bar{R}$ point, as anticipated between the opposite Chern number multi-Weyl nodes.

\section{Electronic structure for $\text{CoGe}$ for space group $C2/m$}
\label{sec:C2/m}
\begin{figure}[ht!]
\centering
\includegraphics[width=\linewidth]{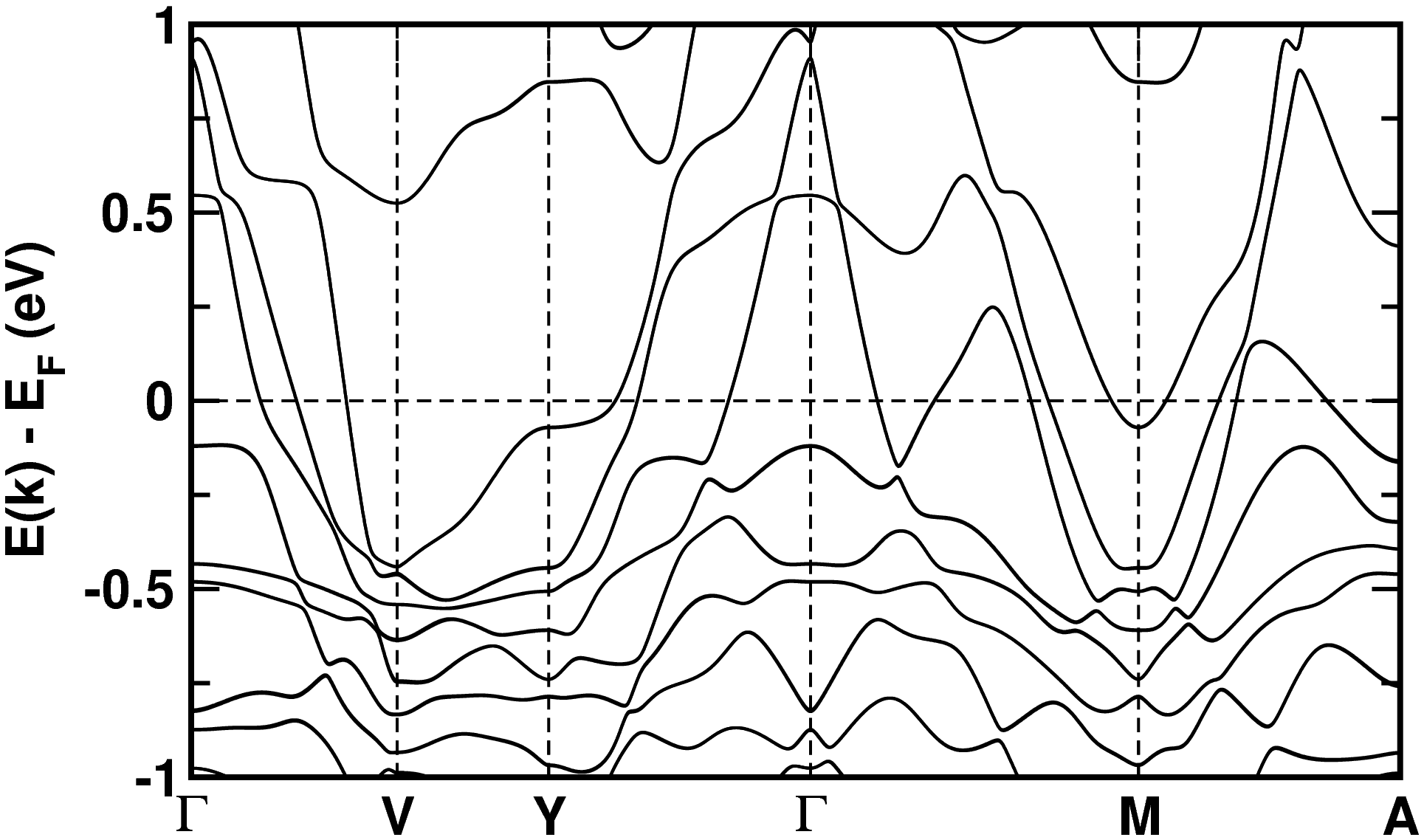}
\caption{(Color online ) Band structure along high symmetry directions for CoGe crystallizes in space group $C2/m$.  }
\label{spg12}
\end{figure}

\begin{figure*}
\includegraphics[width=.7\linewidth]{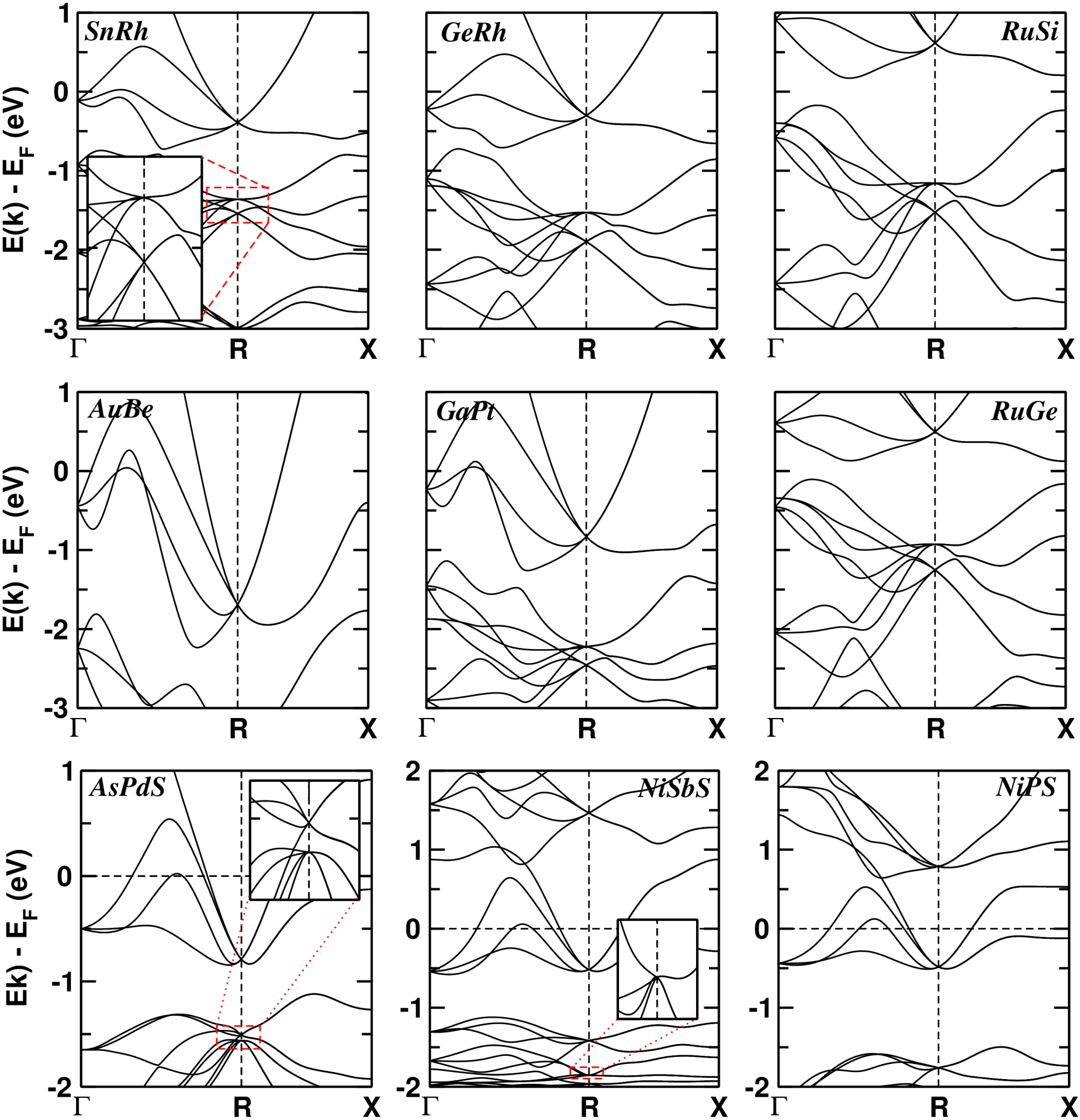}
\caption{(Color online) Band structure for few others binary and ternary systems of space group 198. The inset in the figures are enlarged view of closely spaced four-fold degenerate nodes.  }
\label{fig_bin-tern}
\end{figure*}

As mentioned in the main manuscript, CoGe actually exist in the space group 198 (P2$_1$3)\cite{CoGeExpt} as a metastable phase at ambient condition and high pressure is, apparently, only needed for the synthesis.

However, if the synthesis is carried out at ambient pressure, CoGe crystallizes in the SG 12 ($C2/m$).\cite{CoGeSPG12} All symmetry-protections that hold for SG 198 are not expected to be valid for SG 12. Figure~\ref{spg12} shows the electronic structure of CoGe crystal in space group $C2/m$. Indeed, the band structure clearly shows the absence of the multi-fold band degeneracies at time reversal invariant momenta ($\Gamma$ and $R$).

\begin{figure*}[ht!]
\includegraphics[width=0.9\linewidth]{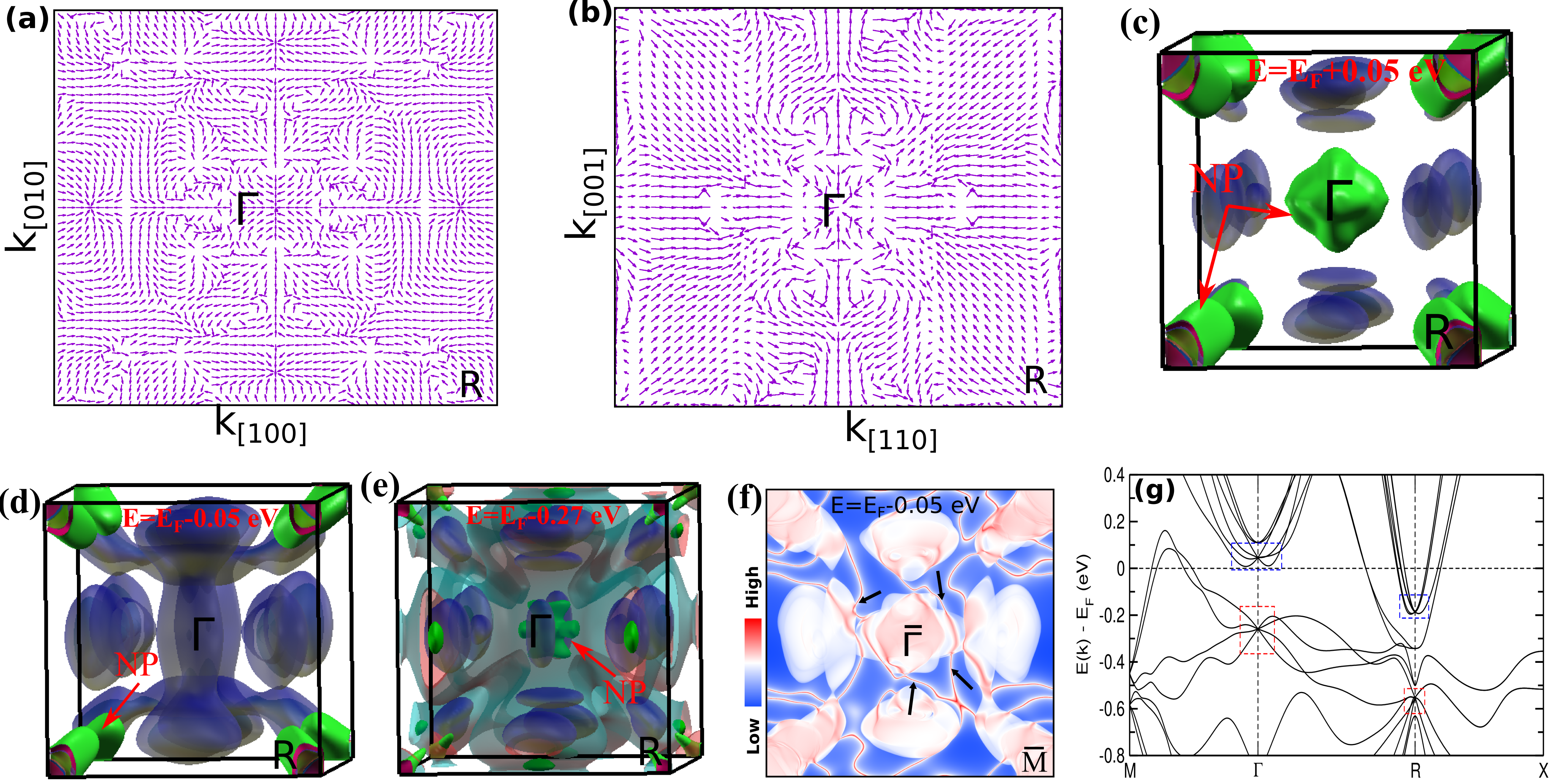}
\caption{(Color online) For BiSbPt with SOI, Berry Curvature plotted in (a) $k_z =0$ and (b) $k_x = k_y$ plane, highlighting its flows between R and $\Gamma$ points in agreement with the sign of the topological charges. (c-e) 3D Fermi surface at three different energy cuts E$_F$+0.05, E$_F$-0.05 and E$_F$-0.27 eV. Fermi pockets corresponding to the non-trivial bands are labeled as NP, and shown by arrowheads. (f) Fermi arc plot at an energy window E$_F$-0.05 eV. (g) Bulk band structure for BiSbPt. Energy location of the multi-fold degenerate Weyl nodes at $\Gamma$ and R points are marked by colored dotted boxes. }
\label{figBiSbPt}
\end{figure*}

\begin{figure*}[ht!]
\includegraphics[width=0.7\linewidth]{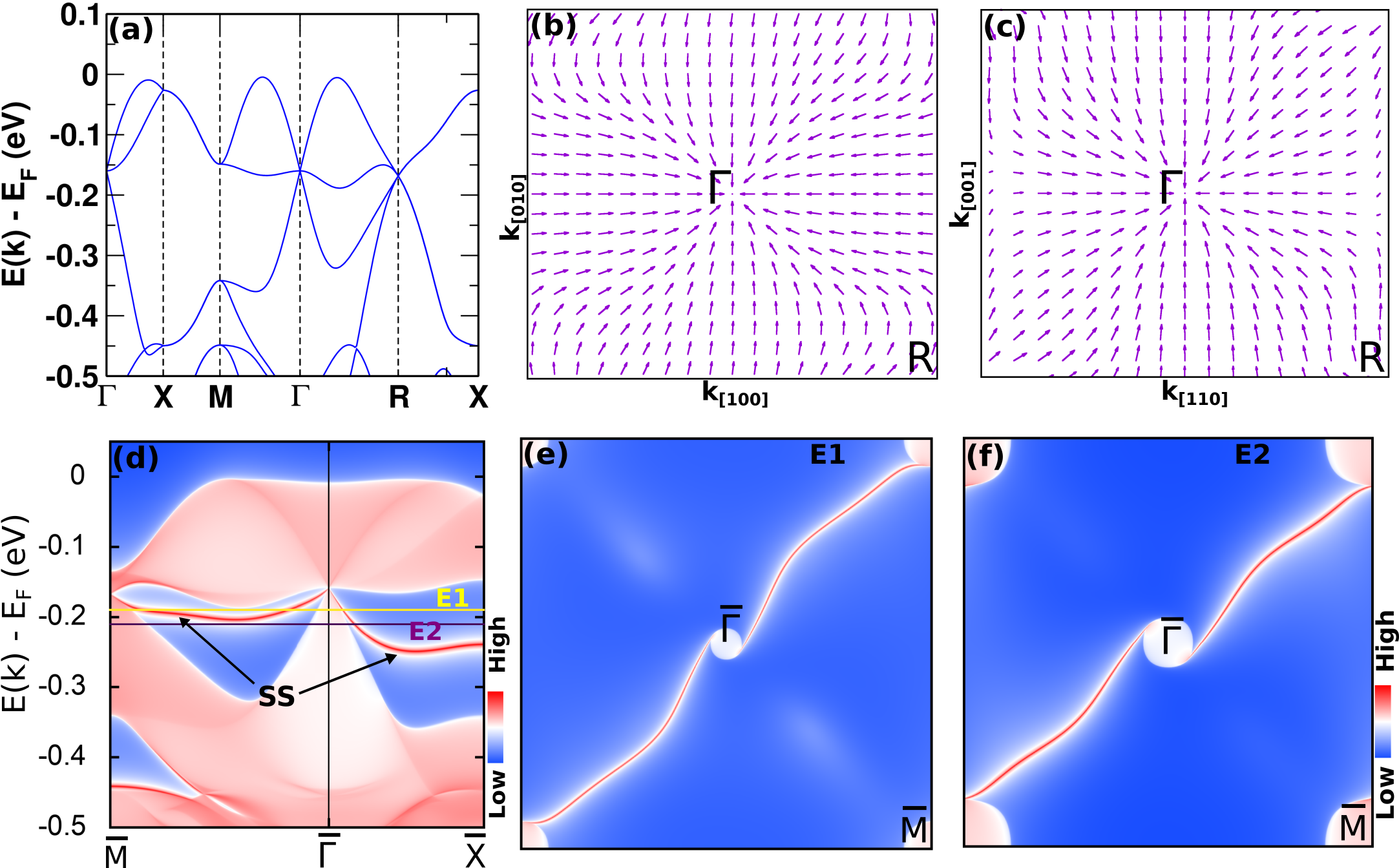}
\caption{(Color online) For KMgBO$_3$ without SOI. (a) Bulk band structure, (b,c) Berry Curvature plotted in k$_z$ =0 (left) and k$_x$ = k$_y$ (right) plane. (d) Surface spectrum at side surface (001). Surface
states are marked by SS. (e,f) Fermi arc contour at energy windows E1=E$_F$-0.19eV and E2=E$_F$-0.21eV, shown by horizontal yellow and purple line in (d).}
\label{figKMgBO-nosoc}
\end{figure*}

\begin{figure*}[ht!]
\includegraphics[width=0.7\linewidth]{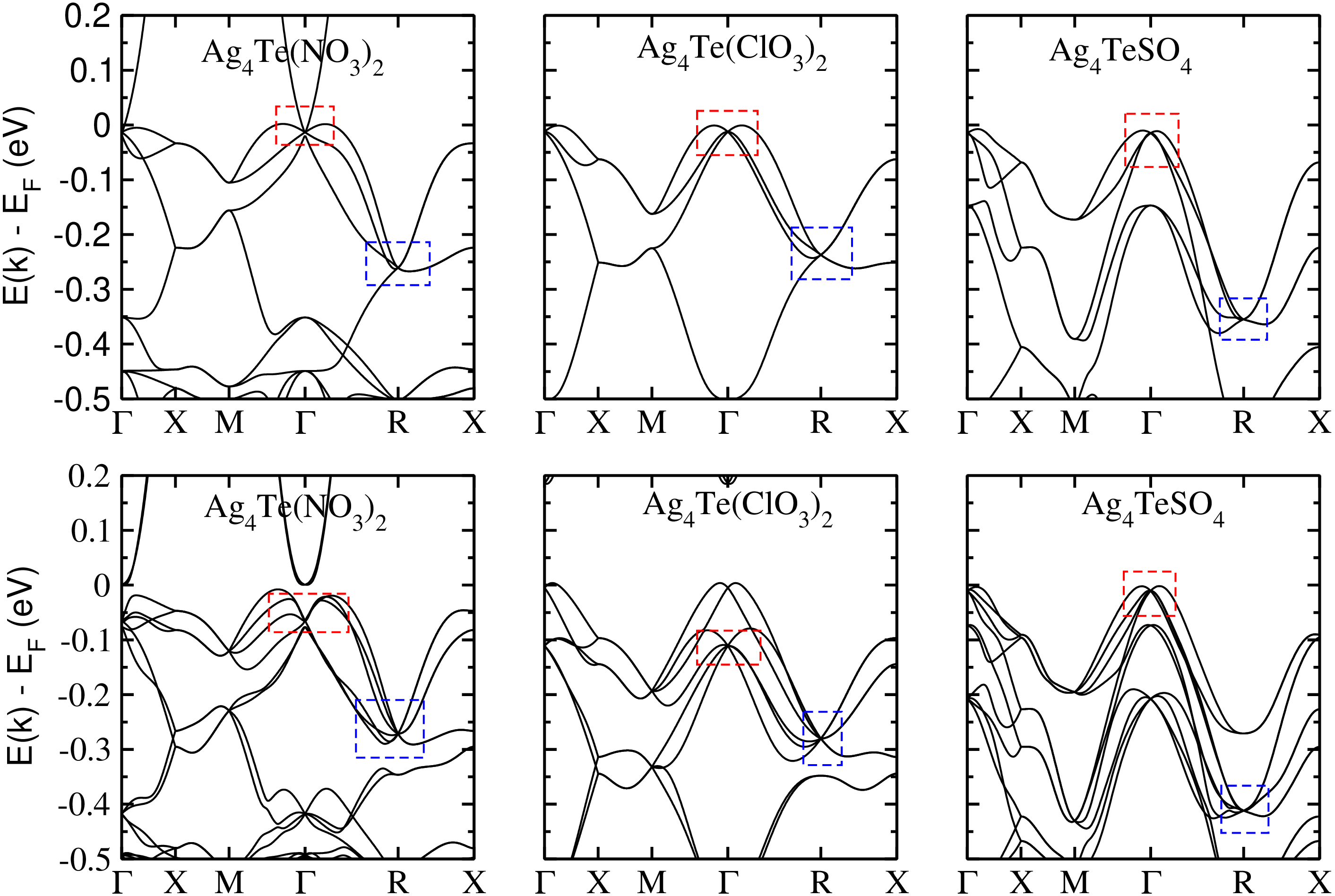}
\caption{(Color online) Band structures for few other quaternary compounds. Top(bottom) panel is without(with) SOI. Spin-1 and Charge-2 Dirac nodes are marked by blue and red dotted boxes in the absence of SOI effect (top panel). In the presence of SOI, four and six-fold degenerate Weyls nodes at $\Gamma$ and R point are marked by red and blue colored boxes (the bottom panel). The multi-fold degenerate Weyl nodes lie at/near the Fermi level. Various degeneracies at the nodal points in this case are protected by non-symmorphic screw rotation, three fold rotation symmetries of SG 198 and time-reversal symmetry. }
\label{figABCD}
\end{figure*}

\section{Four-fold degeneracy at $R$ point for spinless case in few other binary and ternary systems of space group 198 }
\label{subsec:examples}

In addition to CoGe in main article, here in Fig.~\ref{fig_bin-tern}, we showcase the band structures of few other binary and also ternary systems of space group 198 in the absence of spin-orbit coupling. Similar to CoGe, the electronic structure in all these binary and ternary systems shows the four-fold degeneracy at $R$ point irrespective of their location with respect to Fermi level. Thus, the four-fold degeneracy at $R$ point for the spinless case is independent of both the chemical elements at the lattice sites and number of atoms in the cell. Rather, the degeneracy at R point is strictly determined by the crystal space group symmetry.

\section{Berry curvature, Fermi surface and Fermi arc of $\text{BiSbPt}$}
\label{sec:BiSbPt}
In this section, we discuss the Berry curvature, bulk Fermi surfaces and surface Fermi arc properties of a ternary compound BiSbPt.  The Berry curvature on k$_z$=0 and k$_x$=k$_y$ planes are shown in Fig.~\ref{figBiSbPt}(a,b). The flow of Berry curvature from R to $\Gamma$ is in accordance with the existence of Weyl nodes with opposite Chern number at R and $\Gamma$ points as discussed in the main manuscript. Notably, few other nodal sink and source points of Berry flux are also visible on the Berry curvature plots. These points correspond to the spin-1/2 Weyl points which arises due to the accidental band crossings, occurring away from the $\Gamma$ and R points. Figure~\ref{figBiSbPt}(c-e) shows the bulk Fermi surfaces (FSs) of BiSbPt at three different energy isolevel (see caption). Figure~\ref{figBiSbPt}(c) is the FS at isolevel E$_F$+0.05 eV which contain two electron like non-trivial FSs (as indicated by green surface area) at $\Gamma$ and R points. These non-trivial FSs are contributed by the four and six-fold degenerate Weyl nodes marked by blue colored box in Fig.~\ref{figBiSbPt}(g). At a slightly lower energy isolevel of E$_F$-0.05 eV, non-trivial FS at $\Gamma$ point completely disappear and the FS pockets at R point became smaller (see Fig.~\ref{figBiSbPt}(d)). At this energy cut, the non-trivial FSs at R point are dominantly contributed by six-fold degenerate nodal points which is marked by blue colored box at R point in Fig.~\ref{figBiSbPt}(g). At a further lower energy isolevel of E$_F$-0.27 eV, the FSs are majorly contributed by trivial bands, as shown in Fig.~\ref{figBiSbPt}(e). Nevertheless, a tiny FS at zone center is present at this energy cut, as indicated by a red arrow in Fig.~\ref{figBiSbPt}(e). This tiny FS is associated with the four-fold spin-3/2 excitations of the lower node (see red box region in Fig.~\ref{figBiSbPt}(g)) at $\Gamma$ point. 

{\par} Figure~\ref{figBiSbPt}(f) shows the simulated Fermi arc at an energy window 0.05 eV below E$_F$. As expected, four Fermi arcs (indicated by black arrows) emerge from zone center and they propagate towards the corners of the BZ. These four arcs correspond to the upper set of nodal points (inside blue dotted box in Fig.~\ref{figBiSbPt}(g)) near the Fermi level. There are also few other scattered arcs in the plot which are expected to appear from lower energy nodes. Since, the multi-fold Weyl nodes in BiSbPt are surrounded by several other Fermi pockets (as evident from FS maps), the Fermi arcs are somewhat coupled with the trivial Fermi pockets and not completely detached from the bulk band density.

\section{Quaternary alloys}
\label{sec:quaternary}
In addition to binary and ternary compounds, we found few experimentally synthesized\cite{AgTeNO,AgTeSO,KMgBO3-2010} quaternary systems belonging to the space group 198.
Here we provide a detailed calculations for the prototype system KMgBO$_3$ in absence of SOI. Results including SOI effect are given in main manuscript. Figure~\ref{figKMgBO-nosoc}(a) shows the bulk band structure for KMgBO$_3$ at the optimized lattice parameter in absence of SOI. At $\Gamma$ point, we observe the  three-fold degenerate spin-1 Weyl node around $0.15$ eV below the Fermi level (E$_F$). On the other hand, a four-fold degenerate charge-2 nodal point is found at around $0.16$ eV below the E$_F$. These band degeneracies are in accordance with our symmetry analysis for spinless case. We found, Chern number is $\pm2$ for charge-2 nodal point and spin-1 Weyl node at R and $\Gamma$ point respectively. Figure~\ref{figKMgBO-nosoc}(b,c) shows the in-plane Berry curvature ($\vec{\Omega}$) plots on k$_z$=0 and k$_x$=k$_y$ plane. The $\vec{\Omega}$ flows from R to $\Gamma$ point which is in agreement with the computed Chern number. The topological signatures of these unconventional fermions are further studied on the surface. Figure~\ref{figKMgBO-nosoc}(d) shows the surface spectrum for KMgBO$_3$ on (001) surface without including SOI. A pair of surface state (SS) emerged from the $\bar{\Gamma}$ in agreement with the Chern number (C=-2) at $\Gamma$ point. The Fermi arcs (FAs) spectral density are shown in Fig.~\ref{figKMgBO-nosoc}(e,f) at two different energy cuts (see caption). Since in absence of SOI the Chern number at R and $\Gamma$ is $\pm2$, a pair of FAs are nested between $\bar{\Gamma}$ and $\bar{R}$ point of opposite Chern number. Interestingly, the FAs are large and untouched from the bulk states which make KMgBO$_3$ a promising topological semimetal candidate for future photo-emission and transport measurements.

{\par} We showcase bulk band structures for few other quaternaries in Fig.~\ref{figABCD}. The band structures for Ag$_4$Te(NO$_3$)$_2$ and Ag$_4$Te(ClO$_3$)$_2$ are simulated using theoretically optimized lattice parameter at 9.47\AA~and 9.75\AA ~respectively, whereas we used experimental crystal structure for Ag$_4$TeSO$_4$.\cite{AgTeSO} The lower(upper) panel indicates the band structures of these three compounds with (without) SOI effect. Similar to binary and ternary cases, these quaternary compounds also show three-fold (spin-1) and four-fold (charge-2 Dirac) nodes at $\Gamma$ and R point in absence of SOI as shown in Fig.~\ref{figABCD} (upper panel). When SOI is included, symmetry allowed four and six-fold degeneracies are found at $\Gamma$ and R point respectively (see lower panel in Fig.~\ref{figABCD}). The emergence of these degenerate nodal points are consistent with our symmetry analysis and thus reconfirms the validity of our generalized symmetry arguments. As a matter of fact, our symmetry based arguments are independent of the number of constituent elements and lattice sites of the concerned compound, as long as they belong to the SG 198. Most importantly, the Weyl nodes in these quaternaries lie very close to the Fermi level with almost negligible extra trivial band crossings. These makes them potential candidates for future photo-emission experiments.


\onecolumngrid
\newpage 
\section{Compendium of symmetry relations}
\label{sec:derivations}
We recall the various formulas that will be used below.
\begin{align*}
\textcolor{blue}{\{\mathcal{O}_1|\vec{t_1}\} \{\mathcal{O}_2|\vec{t_2}\}} \textcolor{blue}{=}\;& \textcolor{blue}{\{\mathcal{O}_{1}\mathcal{O}_{2}|\mathcal{O}_{1}\vec{t_2}+\vec{t_1}\}} \\
\textcolor{magenta}{\{\mathcal{O}|\vec{t}\}^{-1}} \textcolor{magenta}{=}\;& \textcolor{magenta}{\{\mathcal{O}^{-1}|-\mathcal{O}^{-1}\vec{t}\}} \\
	\textcolor{ForestGreen}{\{\mathbb{I}|\vec{t}\} =}\;& \textcolor{ForestGreen}{e^{-i\vec{k}.\vec{t}}} \\
	\textcolor{violet}{C_{2x}\left(x,y,z \right )\longrightarrow}\;& \textcolor{violet}{\left(x,-y,-z \right ) } \\
	\textcolor{violet}{C_{2y}\left(x,y,z \right )\longrightarrow}\;& \textcolor{violet}{\left(-x,y,-z \right )} \\
	\textcolor{violet}{C_{2z}\left(x,y,z \right )\longrightarrow}\;& \textcolor{violet}{\left(-x,-y,z \right )} \\
	\textcolor{violet}{C_{3,111}\left(x, y, z \right )\longrightarrow}\;&  \textcolor{violet}{\left(z, x, y \right )} \\
	\textcolor{violet}{C^{-1}_{3,111}\left(x, y, z \right )\longrightarrow}\;&  \textcolor{violet}{\left(y, z, x \right )} \\
	\textcolor{ginger}{\left(x_1,y_1,z_1 \right )+\left(x_2,y_2,z_2 \right )\longrightarrow}\;& \textcolor{ginger}{\left(x_1+x_2,y_1+y_2,z_1+z_2\right ) }
\end{align*}
From the first relation, it also follows that
\begin{align*}
\textcolor{cyan}{\{\mathcal{O}_1 \mathcal{O}_2|x,y,z \}} \textcolor{cyan}{=}\;& \textcolor{cyan}{\{\mathcal{O}_1|p,q,r \}\{\mathcal{O}_2|l,m,n \}\{\mathbb{I}|a,b,c \}} \\
\textcolor{cyan}{\{\mathcal{O}|\tilde{u},\tilde{v},\tilde{w}\}} \textcolor{cyan}{=}\;&\textcolor{cyan}{\{\mathcal{O}|u,v,w\}\{\mathbb{I}|\alpha,\beta,\gamma\}} \\
\end{align*}
where $\textcolor{black}{\left(x,y,z\right)} \textcolor{black}{=}\textcolor{black}{ \mathcal{O}_1\left(\mathcal{O}_2 \left(a,b,c\right) + \left(l,m,n\right) \right) +\left(p,q,r\right) }$, $\left(\tilde{u},\tilde{v},\tilde{w}\right)=\mathcal{O}\left(\alpha,\beta,\gamma\right)+\left(u,v,w\right)$ and $\mathbb{I}$ is identity. This relation is also useful in some of the derivations.

The color scheme set up above will be used
in the remaining text when needed to allow for easy parsing of the various algebraic manipulations. In some algebraic manipulation, any expression with a given color in any line is replaced in the following line by the right hand side of the corresponding colored formula above.
{\par} Various algebraic equations given in the Appendix of the main manuscript are explicitly derived below:

\textbf{Eq. A.2} $\mathbf{\left( S^{2}_{2y} = 1 \right)}$:-- 
\begin{align*}
& S^{2}_{2y} \\
=\;& \textcolor{blue}{\{C_{2y}|0,\frac{1}{2},\frac{1}{2}\}\{C_{2y}|0,\frac{1}{2},\frac{1}{2}\}} \nonumber \\
=\;& \{C^{2}_{2y}|\textcolor{violet}{C_{2y}\left(0,\frac{1}{2},\frac{1}{2} \right )}+\left(0,\frac{1}{2},\frac{1}{2} \right )\} \nonumber \\
=\;& \{C^{2}_{2y}|\textcolor{ginger}{\left(0,\frac{1}{2},\bar{\frac{1}{2}} \right ) + \left(0,\frac{1}{2},\frac{1}{2} \right )} \} \nonumber \\
=\;& \{\textcolor{red}{C^{2}_{2y}}|0,1,0\} \nonumber \\
=\;& \{\mathcal{R}|0,1,0\} \nonumber \\
=\;& \{\textcolor{ForestGreen}{\mathbb{I}|0,1,0\}} \nonumber \\
=\;& 1 \\
\Rightarrow \;& S^{2}_{2y} = 1
\end{align*}


\textbf{Eq. A.3} $\mathbf{\left( S^{3}_{3} = 1 \right)}$:-- 

\begin{align*}
& S^{3}_{3} \\
=\;& \textcolor{blue}{\{C_{3,111}|0,0,0\}^{3}} = \{\textcolor{red}{C^{3}_{3,111}}|0,0,0 \} \\
=\;& \{\mathcal{R}|0,0,0 \} \\
=\;& \textcolor{ForestGreen}{\{\mathbb{I}|0,0,0 \}} \\
=\;& 1 \\
\Rightarrow \;& S^{3}_{3} = 1
\end{align*}

\textbf{Eq. A.4a} $\mathbf{\left( \left[S_{2z},S_{2y}\right] = 0 \right)}$:-- 
\begin{align*}
&S_{2z}S_{2y} \\
=& \; \textcolor{blue}{\{C_{2z}|\frac{1}{2},0,\frac{1}{2}\}\{C_{2y}|0,\frac{1}{2},\frac{1}{2}\}} \nonumber \\
=\;& \{C_{2z}C_{2y}| \textcolor{violet}{C_{2z}\left(0,\frac{1}{2},\frac{1}{2} \right )} + \left(\frac{1}{2},0,\frac{1}{2} \right )   \}   \nonumber \\
=\;& \{ C_{2z}C_{2y}|\frac{1}{2},\bar{\frac{1}{2}},1  \} \nonumber \\
=\;&  \{\mathcal{R} C_{2y}C_{2z}|\frac{1}{2},\bar{\frac{1}{2}},1  \} \nonumber \\
=\;&  \{\textcolor{red}{\mathbb{I} C_{2y}}C_{2z}|\frac{1}{2},\bar{\frac{1}{2}},1  \} \\
=\;&  \textcolor{cyan}{\{C_{2y}C_{2z}|\frac{1}{2},\bar{\frac{1}{2}},1  \}} \\
=\;& \{ C_{2y}|0,\frac{1}{2},\frac{1}{2} \} \{ C_{2z}|\frac{1}{2},0,\frac{1}{2} \} \{\mathbb{I}|1,1,\bar{1} \}  \nonumber \\
=\;& S_{2y}S_{2z} \textcolor{ForestGreen}{\{\mathbb{I}|1,1,\bar{1} \}} \nonumber \\
=\;&  S_{2y}S_{2z} \nonumber \\
\Rightarrow &\;	 [ S_{2z},S_{2y}] = 0 
\end{align*}

\textbf{Eq. A.4b} $\mathbf{\left(S_{2z}S_{3} = S_{3}S_{2y} \right)}$:-- 

\begin{align*}
&S^{-1}_{3}S_{2z}S_{3} \\
=& \;  \textcolor{magenta}{\{C_{3,111}|0,0,0\}^{-1}} \{C_{2z}|\frac{1}{2},0,\frac{1}{2}\}\{C_{3,111}|0,0,0\} \nonumber \\
=& \;  \{C^{-1}_{3,111}|0,0,0\} \textcolor{blue}{\{C_{2z}|\frac{1}{2},0,\frac{1}{2}\}\{C_{3,111}|0,0,0\}} \nonumber \\
=\;&  \textcolor{blue}{\{C^{-1}_{3,111}|0,0,0\}\{C_{2z}C_{3,111}|\frac{1}{2},0,\frac{1}{2}\}} \nonumber \\
=\;&  \textcolor{red}{\{ C^{-1}_{3,111}C_{2z}C_{3,111}}|C^{-1}_{3,111}\left(\frac{1}{2},0,\frac{1}{2} \right ) \} \nonumber \\
=\;&  \{C_{2y}|\textcolor{violet}{C^{-1}_{3,111}\left(\frac{1}{2},0,\frac{1}{2} \right )} \} \\
=\;&  \{C_{2y}|0,\frac{1}{2},\frac{1}{2}\}\nonumber \\
=\;&  S_{2y} \nonumber \\
\Rightarrow &\;	 S_{2z}S_{3} =  S_{3}S_{2y}
\end{align*}

\textbf{Eq. A.4c} $\mathbf{\left(S_{3}S_{2z}S_{2y} = S_{2y}S_{3}  \right)}$:-- 
\begin{align*}
& S_{3}S_{2z}S_{2y}S^{-1}_{3} \\
= &\; \{C_{3,111}|0,0,0\}\textcolor{blue}{\{C_{2z}|\frac{1}{2},0,\frac{1}{2}\}}\textcolor{blue}{\{C_{2y}|0,\frac{1}{2},\frac{1}{2}\}}\{C_{3,111}|0,0,0\}^{-1} \\
=\;& \textcolor{blue}{\{C_{3,111}|0,0,0\}\{C_{2z}C_{2y}|\frac{1}{2},\bar{\frac{1}{2}},1\}}\{C_{3,111}|0,0,0\}^{-1} \\
=\;& \{C_{3,111}C_{2z}C_{2y}|1,\frac{1}{2},\bar{\frac{1}{2}}\} \textcolor{magenta}{\{C_{3,111}|0,0,0\}^{-1} } \\
=\;& \textcolor{blue}{\{C_{3,111}C_{2z}C_{2y}|1,\frac{1}{2},\bar{\frac{1}{2}}\} \{C^{-1}_{3,111}|0,0,0\}} \\
=\;& \textcolor{red}{\{C_{3,111}C_{2z}C_{2y}C^{-1}_{3,111}}|1,\frac{1}{2},\bar{\frac{1}{2}}\} \\
=\;& \{C_{2y}|1,\frac{1}{2},\bar{\frac{1}{2}}\} \\
=\;& \{C_{2y}|0,\frac{1}{2},\frac{1}{2}\} \textcolor{ForestGreen}{\{\mathbb{I}|\bar{1},0,1\}} \\
=\;& S_{2y} \\
\Rightarrow \;& S_{3}S_{2z}S_{2y} =  S_{2y}S_{3} 
\end{align*}

\textbf{Eq. A.7a} $\mathbf{\left(S^{2}_{2x} = -1  \right)}$:-- 
\begin{align*}
& S^{2}_{2x} \\
= &\;\textcolor{blue}{\{C_{2x}|\frac{1}{2},\frac{3}{2},0\}\{C_{2x}|\frac{1}{2},\frac{3}{2},0\}} \nonumber \\
=\;& \{C^{2}_{2x}| \textcolor{violet}{C_{2x}\left(\frac{1}{2},\frac{3}{2},0 \right )}+\left(\frac{1}{2},\frac{3}{2},0 \right )\} \nonumber \\
=\;& \{C^{2}_{2x}|\textcolor{ginger}{\left(\frac{1}{2},\bar{\frac{3}{2}},0 \right )+\left(\frac{1}{2},\frac{3}{2},0 \right )}\} \nonumber \\
=\;& \{\textcolor{red}{C^{2}_{2x}}|1,0,0\} \nonumber \\
=\;& \{\mathcal{R}|1,0,0 \} \nonumber \\
=\;& \textcolor{ForestGreen}{\{\mathbb{I}|1,0,0 \}} \nonumber \\
=\;& e^{-i\pi} \nonumber \\
=\;&  -1 \\
\Rightarrow \;& S^{2}_{2x}= -1
\end{align*}

\textbf{Eq. A.7b} $\mathbf{\left( S^{2}_{2y} = -1  \right)}$:-- 
\begin{align*}
 & S^{2}_{2y} \\
 =\;& \textcolor{blue}{ \{C_{2y}|0,\frac{3}{2},\frac{1}{2}\} \{C_{2y}|0,\frac{3}{2},\frac{1}{2}\}} \nonumber \\
 =&\; \{C^{2}_{2y}|\textcolor{violet}{ C_{2y}\left(0,\frac{3}{2},\frac{1}{2} \right )}+\left(0,\frac{3}{2},\frac{1}{2} \right )\} \nonumber \\
 =&\; \{C^{2}_{2y}|\textcolor{ginger}{\left(0,\frac{3}{2},\bar{\frac{1}{2}} \right )+\left(0,\frac{3}{2},\frac{1}{2} \right )}\}  \nonumber \\
 =&\; \{\textcolor{red}{C^{2}_{2y}}|0,3,0\} \nonumber \\
 =&\; \{\mathcal{R}|0,3,0 \} \nonumber \\
 =&\; \textcolor{ForestGreen}{\{\mathbb{I}|0,3,0\}} \nonumber \\
 =&\; e^{-3i\pi} \nonumber \\
 =&\;  -1 \\
 \Rightarrow \;& S^{2}_{2y} = -1
\end{align*}

\textbf{Eq. A.7c} $\mathbf{\left( S_{2x}S_{2y} = -S_{2y} S_{2x}   \right)}$:-- 
\begin{align*}
&S_{2x}S_{2y} \\
=&\; \textcolor{blue}{ \{C_{2x}|\frac{1}{2},\frac{3}{2},0\} \{C_{2y}|0,\frac{3}{2},\frac{1}{2}\}} \nonumber \\
=\;& \{C_{2x}C_{2y}|\frac{1}{2},0,\bar{\frac{1}{2}}\}  \nonumber \\
=\;& \{\mathcal{R}C_{2y}C_{2x}|\frac{1}{2},0,\bar{\frac{1}{2}}\} \nonumber \\
=\;& \{\textcolor{red}{\mathbb{I}C_{2y}}C_{2x}|\frac{1}{2},0,\bar{\frac{1}{2}}\} \nonumber \\
=\;& \textcolor{cyan}{\{C_{2y}C_{2x}|\frac{1}{2},0,\bar{\frac{1}{2}}\} }\nonumber \\
=\;& \{C_{2y}|0,\frac{3}{2},\frac{1}{2}\} \{C_{2x}|\frac{1}{2},\frac{3}{2},0 \} \{\mathbb{I}|\bar{1},3,\bar{1} \} \nonumber \\
=\;& S_{2y}S_{2x}  \textcolor{ForestGreen}{ \{\mathbb{I}|\bar{1},3,\bar{1} \} } \nonumber \\
=\;&  S_{2y} S_{2x} e^{-i\left(-\pi+3\pi-\pi \right )}  \nonumber \\
=\;&  -S_{2y} S_{2x}  \nonumber \\
\Rightarrow \;&  S_{2x}S_{2y} = -S_{2y} S_{2x}
\end{align*}

\textbf{Eq. A.7d} $\mathbf{\left( S_{2x}S_{3} = S_{3} S_{2y}  \right)}$:-- 
\begin{align*}
& S_{3}^{-1}S_{2x}S_{3} \\
=&\; \textcolor{magenta}{ \{C^{-1}_{3,111}|0,1,0\}^{-1}} \{C_{2x}|\frac{1}{2},\frac{3}{2},0\} \{C^{-1}_{3,111}|0,1,0\}  \\
=\;&  \{C_{3,111}|- \textcolor{violet}{ C_{3,111}\left(0,1,0 \right )\}} \{C_{2x}|\frac{1}{2},\frac{3}{2},0\}  \{C^{-1}_{3,111}|0,1,0\}  \\
=\;&  \textcolor{blue}{ \{C_{3,111}|0,0,-1\} \{C_{2x}|\frac{1}{2},\frac{3}{2},0\}} \{C^{-1}_{3,111}|0,1,0\}  \\
=\;&  \{C_{3,111}C_{2x}| \textcolor{violet}{ C_{3,111}\left(\frac{1}{2},\frac{3}{2},0 \right )}+\left(0,0,-1 \right )\}  \{C^{-1}_{3,111}|0,1,0\}   \\
=\;&  \{C_{3,111}C_{2x}|\textcolor{ginger}{\left(0,\frac{1}{2},\frac{3}{2} \right )+\left(0,0,-1 \right )}\} \{C^{-1}_{3,111}|0,1,0\}  \\
=\;&  \textcolor{blue}{ \{C_{3,111}C_{2x}|0,\frac{1}{2},\frac{1}{2}\} \{C^{-1}_{3,111}|0,1,0\} }   \\
=\;&  \{C_{3,111}C_{2x}C^{-1}_{3,111}| \textcolor{violet}{ C_{3,111}C_{2x}\left(0,1,0 \right )}+\left(0,\frac{1}{2},\frac{1}{2} \right )\} \\ 
=\;&  \{C_{3,111}C_{2x}C^{-1}_{3,111}| \textcolor{ginger}{ \left(0,0,\bar{1} \right )+\left(0,\frac{1}{2},\frac{1}{2} \right )}\} \\ 
=\;&  \{ \textcolor{red}{ C_{3,111}C_{2x}C^{-1}_{3,111}}|0,\frac{1}{2},\bar{\frac{1}{2}}\}  \\
=\;&  \{C_{2y}|0,\frac{3}{2},\frac{1}{2}\} \textcolor{ForestGreen}{\{\mathbb{I}|0,\bar{1},1\}} \\
=\;& \{C_{2y}|0,\frac{3}{2},\frac{1}{2}\} e^{-i\left(-\pi+\pi \right )}  \\
=\;&  S_{2y} \\
\Rightarrow \;&  S_{2x}S_{3} = S_{3} S_{2y}
\end{align*}

\textbf{Eq. A.7e} $\mathbf{\left( S_{3}S_{2x}S_{2y} =  S_{2y}S_{3}  \right)}$:-- 
\begin{align*}
& S_{3}S_{2x}S_{2y}S_{3}^{-1} \\
=&\; \{C^{-1}_{3,111}|0,1,0\}  \textcolor{blue}{ \{C_{2x}|\frac{1}{2},\frac{3}{2},0\}\{C_{2y}|0,\frac{3}{2},\frac{1}{2}\}} \\ & \{C^{-1}_{3,111}|0,1,0\}^{-1}  \\
=\;&  \textcolor{blue}{ \{C^{-1}_{3,111}|0,1,0\} \{C_{2x}C_{2y}|\frac{1}{2},0\bar{\frac{1}{2}}\} }  \{C^{-1}_{3,111}|0,1,0\}^{-1}  \\
=\;&  \{C^{-1}_{3,111}C_{2x}C_{2y}| \textcolor{violet}{ C^{-1}_{3,111}\left(\frac{1}{2},0\bar{\frac{1}{2}} \right )}+\left(0,1,0 \right )\} \\&  \textcolor{magenta}{ \{C^{-1}_{3,111}|0,1,0\}^{-1}}  \\ 
=\;&  \{C^{-1}_{3,111}C_{2x}C_{2y}|0,\frac{1}{2},\frac{1}{2}\} \{C_{3,111}|- \textcolor{violet}{ C_{3,111}\left(0,1,0 \right )}  \}  \\
\end{align*}
\begin{align*}
=\;&  \textcolor{blue}{ \{C^{-1}_{3,111}C_{2x}C_{2y}|0,\frac{1}{2},\frac{1}{2}\} \{C_{3,111}|0,0,\bar{1}\}}  \\
=\;& \{C^{-1}_{3,111}C_{2x}C_{2y}C_{3,111}| \textcolor{violet}{ C^{-1}_{3,111}C_{2x}C_{2y}\left(0,0,\bar{1} \right )} +\left(0,\frac{1}{2},\frac{1}{2} \right )\}  \\
=\;& \{C^{-1}_{3,111}C_{2x}C_{2y}C_{3,111}| \textcolor{ginger}{\left(0,\bar{1},0 \right ) +\left(0,\frac{1}{2},\frac{1}{2} \right )}\}  \\
=\;& \{\textcolor{red}{ C^{-1}_{3,111}C_{2x}C_{2y}C_{3,111}}| 0,\bar{\frac{1}{2}},\frac{1}{2} \}  \\
=\;& \textcolor{cyan}{\{C_{2y}|0,\bar{\frac{1}{2}},\frac{1}{2} \}} \\
=\;&  \{C_{2y}|0,\frac{3}{2},\frac{1}{2}\} \textcolor{ForestGreen}{ \{\mathbb{I}|0,\bar{2},0\}}  \\
=\;&  S_{2y} \\
\Rightarrow \;& S_{3}S_{2x}S_{2y} =  S_{2y}S_{3}
\end{align*}

\textbf{Eq. A.8a} $\mathbf{\left( S^{2}_{2z} = -1  \right)}$:-- 
\begin{align*}
& S^{2}_{2z} \\
=&\;  \textcolor{blue}{ \{C_{2z}|\frac{1}{2},0,\frac{1}{2}\}\{C_{2z}|\frac{1}{2},0,\frac{1}{2}\}} \nonumber \\
= \;& \{C^{2}_{2z}| \textcolor{violet}{ C_{2z}\left(\frac{1}{2},0,\frac{1}{2} \right )}+\left(\frac{1}{2},0,\frac{1}{2} \right )\}  \nonumber \\
= \;& \{C^{2}_{2z}|\textcolor{ginger}{\left(\bar{\frac{1}{2}},0,\frac{1}{2} \right )+\left(\frac{1}{2},0,\frac{1}{2} \right )}\} \nonumber \\
= \;& \{\textcolor{red}{C^{2}_{2z}}|0,0,1\} \nonumber \\
= \;& \{\mathcal{R}|0,0,1\} \nonumber \\
= \;& -\textcolor{ForestGreen}{\{\mathbb{I}|0,0,1\}} \nonumber \\
= \;& -1
\end{align*}

\textbf{Eq. A.8b} $\mathbf{\left( S^{2}_{2y} = -1  \right)}$:-- 
\begin{align*}
& S^{2}_{2y} \\
=&\; \textcolor{blue}{ \{C_{2y}|0,\frac{1}{2},\frac{1}{2}\}\{C_{2y}|0,\frac{1}{2},\frac{1}{2}\}} \nonumber \\
= \;& \{C^{2}_{2y}|\textcolor{violet}{ C_{2y}\left(0,\frac{1}{2},\frac{1}{2} \right )}+\left(0,\frac{1}{2},\frac{1}{2} \right )\} \nonumber \\
= \;& \{C^{2}_{2y}|\textcolor{ginger}{\left(0,\frac{1}{2},\bar{\frac{1}{2}} \right ) + \left(0,\frac{1}{2},\frac{1}{2} \right )} \} \nonumber \\
= \;& \{\textcolor{red}{C^{2}_{2y}}|0,1,0\} \nonumber \\
= \;& \{\mathcal{R}|0,1,0\} \nonumber \\
= \;& -\textcolor{ForestGreen}{\{\mathbb{I}|0,1,0\}} \nonumber \\
= \;& -1
\end{align*}

\textbf{Eq. A.8c} $\mathbf{\left( S^{3}_{3} = -1 \right)}$:-- 
\begin{align*}
& S^{3}_{3} \\
=\;& \textcolor{blue}{\{C_{3,111}|0,0,0\}^{3}} = \{\textcolor{red}{C^{3}_{3,111}}|0,0,0 \} \\
=\;& \{\mathcal{R}|0,0,0 \} \\
=\;& -\textcolor{ForestGreen}{\{\mathbb{I}|0,0,0 \}} \\
=\;& -1 \\
\Rightarrow \;& S^{3}_{3} = -1
\end{align*}

\textbf{Eq. A.9a} $\mathbf{\left( S_{2z}S_{2y} = -S_{2y}S_{2z}  \right)}$:-- 
\begin{align*}
& S_{2z}S_{2y} \\
= &\; \textcolor{blue}{ \{C_{2z}|\frac{1}{2},0,\frac{1}{2}\}\{C_{2y}|0,\frac{1}{2},\frac{1}{2}\}} \nonumber \\
= \;& \{C_{2z}C_{2y}|\textcolor{violet}{ C_{2z}\left(0,\frac{1}{2},\frac{1}{2}\} \right )} + \left(\frac{1}{2},0,\frac{1}{2} \right )   \}   \nonumber \\
= \;& \{ C_{2z}C_{2y}|\frac{1}{2},\bar{\frac{1}{2}},1  \} \nonumber \\
= \;& \{\mathcal{R}C_{2y}C_{2z}|\frac{1}{2},\bar{\frac{1}{2}},1  \} \nonumber \\
= \;& \{\textcolor{red}{-\mathbb{I}C_{2y}}C_{2z}|\frac{1}{2},\bar{\frac{1}{2}},1  \} \nonumber \\
= \;& -\textcolor{cyan}{\{C_{2y}C_{2z}|\frac{1}{2},\bar{\frac{1}{2}},1  \}} \nonumber \\
= \;& -\{ C_{2y}|0,\frac{1}{2},\frac{1}{2} \} \{ C_{2z}|\frac{1}{2},0,\frac{1}{2} \} \textcolor{ForestGreen}{\{\mathbb{I}|1,1,\bar{1} \}}  \nonumber \\
= \;& -S_{2y}S_{2z} \nonumber \\
\Rightarrow \;& S_{2z}S_{2y} = -S_{2y}S_{2z} 
\end{align*}

\textbf{Eq. A.9b} $\mathbf{\left( S_{2z}S_{3} = S_{3}S_{2y}  \right)}$:-- 
\begin{align*}
& S^{-1}_{3}S_{2z}S_{3} \\
=&\; \{C_{3,111}|0,0,0\}^{-1}\textcolor{blue}{ \{C_{2z}|\frac{1}{2},0,\frac{1}{2}\}\{C_{3,111}|0,0,0\}} \nonumber \\
=\;& \textcolor{magenta}{ \{C_{3,111}|0,0,0\}^{-1}} \{C_{2z}C_{3,111}|\frac{1}{2},0,\frac{1}{2}\} \nonumber \\
= \;& \textcolor{blue}{ \{C^{-1}_{3,111}|0,0,0\}\{C_{2z}C_{3,111}|\frac{1}{2},0,\frac{1}{2}\}} \nonumber \\
= \;& \{  C^{-1}_{3,111}C_{2z}C_{3,111}|\textcolor{violet}{ C^{-1}_{3,111}\left(\frac{1}{2},0,\frac{1}{2} \right )} + \left( 0,0,0 \right) \} \nonumber \\
= \;& \{ \textcolor{red}{ C^{-1}_{3,111}C_{2z}C_{3,111}}|0,\frac{1}{2},\frac{1}{2}\}\} \nonumber \\
= \;& \{C_{2y}|0,\frac{1}{2},\frac{1}{2}\}\nonumber \\
= \;& S_{2y} \nonumber \\
\Rightarrow \;& S_{2z}S_{3} =  S_{3}S_{2y}
\end{align*}

\textbf{Eq. A.9c} $\mathbf{\left(S_{3}S_{2z}S_{2y} = S_{2y}S_{3}   \right)}$:--
\begin{align*}
& S_{3}S_{2z}S_{2y}S^{-1}_{3} \\
= &\; \{C_{3,111}|0,0,0\} \textcolor{blue}{ \{C_{2z}|\frac{1}{2},0,\frac{1}{2}\}} \textcolor{blue}{ \{C_{2y}|0,\frac{1}{2},\frac{1}{2}\}} \\ & \{C_{3,111}|0,0,0\}^{-1} \\
= &\; \{C_{3,111}|0,0,0\} \{C_{2z}C_{2y}|\textcolor{violet}{C_{2z}\left(0,\frac{1}{2},\frac{1}{2}\right)}+\left(\frac{1}{2},0,\frac{1}{2}\right)\} \\& \{C_{3,111}|0,0,0\}^{-1} \\
\end{align*}
\begin{align*}
= &\; \{C_{3,111}|0,0,0\} \{C_{2z}C_{2y}|\textcolor{ginger}{\left(0,\bar{\frac{1}{2}},\frac{1}{2}\right)+\left(\frac{1}{2},0,\frac{1}{2}\right)\}} \\& \{C_{3,111}|0,0,0\}^{-1} \\
= \;& \textcolor{blue}{ \{C_{3,111}|0,0,0\}\{C_{2z}C_{2y}|\frac{1}{2},\bar{\frac{1}{2}},1\} } \{C_{3,111}|0,0,0\}^{-1} \\
= \;&  \{C_{3,111}C_{2z}C_{2y}|\textcolor{violet}{C_{3,111}\left(\frac{1}{2},\bar{\frac{1}{2}},1\right)}\}  \{C_{3,111}|0,0,0\}^{-1} \\
= \;&  \{C_{3,111}C_{2z}C_{2y}|1,\frac{1}{2},\bar{\frac{1}{2}}\} \textcolor{magenta}{\{C_{3,111}|0,0,0\}^{-1}} \\
= \;& \textcolor{blue}{ \{C_{3,111}C_{2z}C_{2y}|1,\frac{1}{2},\bar{\frac{1}{2}}\} \{C^{-1}_{3,111}|0,0,0\}} \\
= \;& \textcolor{red}{ \{C_{3,111}C_{2z}C_{2y}C^{-1}_{3,111}}|1,\frac{1}{2},\bar{\frac{1}{2}}\} \\
= \;& \textcolor{cyan}{\{C_{2y}|1,\frac{1}{2},\bar{\frac{1}{2}}\}} \\
= \;& \{C_{2y}|0,\frac{1}{2},\frac{1}{2}\} \textcolor{ForestGreen}{ \{\mathbb{I}|\bar{1},0,1\}} \\
= \;& S_{2y} \\
\Rightarrow \;& S_{3}S_{2z}S_{2y} = S_{2y}S_{3} 
\end{align*}
